\documentclass[fleqn,usenatbib]{mnras}

\usepackage{newtxtext,newtxmath}

\usepackage[T1]{fontenc}

\DeclareRobustCommand{\VAN}[3]{#2}
\let\VANthebibliography\thebibliography
\def\thebibliography{\DeclareRobustCommand{\VAN}[3]{##3}\VANthebibliography}

\usepackage{graphicx}	
\usepackage{amsmath}	

\newcommand\nbR{\ensuremath{\mathrm{I\! R}}}
\newcommand\nbE{\ensuremath{\mathrm{I\! E}}}

\title[Clustering galaxy images]{Machine Learning and galaxy morphology: for what purpose?}

   \author[D. Fraix-Burnet]{D. Fraix-Burnet$^1$\thanks{didier.fraix-burnet@univ-grenoble-alpes.fr}
\\
$^{1}$Univ. Grenoble Alpes, CNRS, IPAG, Grenoble, France
}

\date{Received May 15, 2023; accepted }

\pubyear{2023}

\begin{document}
	\label{firstpage}
	\pagerange{\pageref{firstpage}--\pageref{lastpage}}
	\maketitle
	
	\begin{abstract}
Classification of galaxies is traditionally associated with their morphologies through visual inspection of images. The amount of data to come renders this task inhuman and Machine Learning (mainly Deep Learning) has been called to the rescue for more than a decade. However, the results look mitigate and there seems to be a shift away from the paradigm of the traditional morphological classification of galaxies.
In this paper, I want to show that the algorithms indeed are very sensitive to the features present in images, features that do not necessarily correspond to the Hubble or de Vaucouleurs vision of a galaxy.  However, this does not preclude to get the correct insights into the physics of galaxies.
I have applied a state-of-the-art "traditional" Machine Learning clustering tool, called Fisher-EM, a latent discriminant subspace Gaussian Mixture Model algorithm, to 4458 galaxies carefully classified into 18 {types} by the EFIGI project.
The optimum number of clusters given by the Integrated Complete Likelihood criterion is 47. The correspondence with the EFIGI classification is correct, but it appears that the Fisher-EM algorithm gives a great importance to the distribution of light which translates to characteristics such as the bulge to disk ratio, the inclination or the presence of foreground stars. 
The discrimination of some physical parameters (bulge-to-total luminosity ratio, $(B-V)_T$, intrinsic diameter, presence of flocculence or dust, arm strength)  is very comparable in the two classifications. 
	\end{abstract}
	
	
	

\section{Introduction}
\label{Introduction}


The original Hubble morphological classification \citep{Hubble1926} has four categories: ellipticals, spirals, barred spirals and irregulars. Later, more sophisticated classifications were designed. \citet{Schutter2015} provides a thorough history of morphological classification with the detailed nomenclatures that have been introduced. All these classifications come from visual inspection of images in the visible domain. Apart from the original Hubble scheme, the {best known} classification is the {numerical T-type imagined by \citet{deVaucouleurs1963} and used in the RC3 catalog with 18 {stages} \citep[e.g.][]{Baillard2011}}.

It has long been recognised that huge surveys require an automatic classification of the images of galaxies. Machine Learning generates a lot of hopes and in particular the supervised Deep Learning approach seems appropriate. Unfortunately, this requires large and reliable training samples. Astronomers being too few, the idea of citizen science emerged. 

Galaxy Zoo 1 \citep{Lintott2008} considered six classes (ellipticals, spiral clockwise and anticlockwise, edge-on spirals, merger, and star/artefact) that can be easily identified by eye from non experts. Following the success of this first attempt, the Galaxy Zoo 2 project \citep{Willett2013} went further and proposed a decision tree with 11 tasks leading to the identification of 37 features. It is thus more detailed than the de Vaucouleurs classification but noticeably departs from the traditional way to classify galaxies. Actually, the Galaxy Zoo 2 project proposes a \emph{description} of more than 300~000 images of galaxies, with a median of 44 citizen-classifications per galaxy. It seems difficult to extend this kind of work to build a larger training sample suited for the billions of images that will soon populate the data bases \citep{Fielding2022} unless we call on myriad ‘micro-workers’ that support Artificial Intelligence in other domains  \citep[e.g.][]{Tubaro2020}.

In the EFIGI project \citep{Baillard2011}, a small group of astronomers performed the visual classification of 4458 galaxies {into 18 types, very close to the 18 stages of de Vaucouleurs}. One of the aims was to provide a training base with a more refined classification than the six categories of the Galaxy Zoo 1 project. This sample is probably too small for this purpose {but there is no better one up to now. In addition} there is a non-negligible dispersion in the final classification because of disagreements between the expert astronomers. {Single-astronomer classifications would have no dispersion but could yield different results.}

\citet{Buta2019}  made a re-examination of the EFIGI sample performing the classification three times at 6-month intervals. He found some disagreement between the three runs {but on average an excellent agreement with the EFIGI classification}. He concludes that such very detailed classifications might be quite challenging as training samples and are probably more useful at recognizing peculiar structures of interest for the physics and evolution of galaxies.

Using supervised Machine Learning, it appears that classifying into a low number of classes (such as spiral/elliptical/irregular or star/galaxy) is relatively easy for most methods, but a higher number of classes proves very difficult to recover \citep{Polsterer2012}. 
This does not seem to be only due to a lack of a good training sample, and no supervised classification studies have attempted to retrieve the de Vaucouleurs classification.

For instance, \citet{Ma2022} considers five categories: completely round, in-between or cigar-shaped smooth, edge-on and spiral. Interestingly, they propose a technique to take into account the fuzzy nature of the visual classification. 

\citet{Khramtsov2022} performed two analyses, one with the five classes above and one with 34 of the 37 very detailed features (smooth, feature or disk, star or artifact, number of arms, edge-on or not, bar or not...) as defined in the Galaxy Zoo 2 project \citep{Willett2013}. The accuracy obtained for the different features is good to very good (69 to 99\%), and somewhat less so for the five classes. They point out several limitations in the use of Deep Learning on galaxy images and describe specific features (such as the bars) that are {prone to misclassifications}.

\citet{Gharat2022} note that six classes are not sufficient to describe the details visible in the images and define arbitrarily ten classes: face-on disk (no, tight, medium or loose spiral), edge-on disk (no, rounded or boxy bulge), smooth (completely or in-between round, cigar shaped). 

Indeed, supervised studies {appear to focus on the detection of specified features present in the images rather on a given morphological classification. In fact, these }features are often not identified in the de Vaucouleurs scheme. We must be aware that automatic classification cannot be expected to reproduce the detailed visual classification because of its subjectivity. For instance, the EFIGI paper \citep{Baillard2011} provides a precise definition of the {types/stages}. It appears that some criteria are relative: ellipticals are described as having \emph{smooth} intensity distribution with \emph{relatively} steep gradient, are \emph{more or less} elongated, spirals \emph{may} harbour a bar, Sa galaxies have a \emph{low} amounts of dust, Sc galaxies have \emph{fairly} weak {bulges}, or dwarf spheroid elliptical \emph{may} contain a tight nucleus. This is also true for the Galaxy Zoo projects where many in-between situation makes the choice essentially random. It is hard to imagine how an algorithm, be it a sophisticated neural network, can cope with these very subjective properties.

Detecting {unspecified} features in the images can be a more objective task and certainly better matches what is really present and what an algorithm could really detect. As shown above, there seems to be a trend in the literature to "abandon" the traditional morphological classification and rather concentrate on automatic description of features. However, the only training sample available, the Galaxy Zoo 2, is limited in size, depends on visual inspection with the associated human biases.

This is probably why unsupervised learning gained some more interest in the recent years. The goal is to construct objective training samples by letting the machine determine its own descriptors that  can be used to identify the groupings. It is also easy to repeat when new data become available, or {when other information is} necessary to fully characterize galaxies such as spectroscopy or multi-wavelength photometry from the X-ray to the radio.

{A hybrid} approach is proposed by \citet{Schutter2015} for the EFIGI sample and the png composite images of the three bands $r,g$ and $i$ that has been used for the visual classification \citep{Baillard2011}. For each morphological type, they derive descriptors from different transforms of the images such as texture and high-contrast features, polynomial decomposition and pixel statistics.  They claim to perform an unsupervised analysis of galaxy morphologies and their goal is "not to automatically classify galaxies by their morphology, but to quantitatively deduce a network of similarities between the different morphological types using merely the galaxy images". In doing so, they perform a \emph{description} of each class in the space of descriptors, but not an unsupervised \emph{classification} because they need a training phase to associate some descriptors to each of the morphological type as annotated by the EFIGI catalogue. They keep the most discriminant of these descriptors for a subsequent supervised classification and a phylogenetic representation of the de Vaucouleurs classification. {This procedure creates somehow a circular argument that unsurprisingly results in the perfect agreement of the similarity sequence they find with the one imagined by de Vaucouleurs.} We think it should have been possible to perform a truly unsupervised classification from the descriptors themselves.

A truly unsupervised learning procedure consists in detecting structures in the data space that can be identified as clusters or groups and defined as classes. Most often, the sample of images must be projected into a representation space of lower dimensionality in which these clusters can emerge. Unsupervised classification is an exploratory process so that by definition the number of clusters is unknown. Not all approaches propose an objective way to estimate this number.

\citet{Hocking2018} used an unsupervised patch-based model technique through a somewhat complicated procedure to extract sub-image patches and compare the images from these patches. Unfortunately, there is no obvious criterion to decide on the number of clusters, so that some visual inspection may be required. In their paper, this number varies from 2 to 200 depending on the sample. Comparing with the Galaxy Zoo results, they conclude that "there is not a direct mapping between Galaxy Zoo and our hierarchical labelling", but "a good level of concordance between human and machine classifications".  \cite{Martin2020} used the same algorithm on a different sample of about 90~000 images. They arbitrarily chose 160 clusters and described their characteristics by eye. They find a good separation between ellipticals and spirals.

\citet{Cheng2021} used a Variational Auto-Encoder on the EFIGI images, before applying hierarchical clustering. Several choices lead the machine to find a number of 27 clusters. They did not attempt to compare their 27 classes to the 18 de Vaucouleurs {stages}, but instead to a mixture of eight morphological classes. Despite this restriction and despite a very sophisticated and specific procedure adapted to the goal of the study, none of the classes is found to correspond to a pure morphological class. 
Due to the "vagueness of the visual classification" of the classic Hubble types (see above), they conclude that it is not so important that "our unsupervised method matches visual classifications and physical properties, but that the method provides an independent classification that may be more physically meaningful".

In a similar approach, \citet{Fielding2022} used a convolutional encoder, acting as a dimensionality reduction, before performing a clustering with k-means, fuzzy C-means and agglomerative tools. Their goal was to compare the feature detection by these algorithms with respect to the questions asked in the Galaxy Zoo DECaLS project \citep{Walmsley2022} regarding ten features. 
For each question/feature, the number of clusters to be given as input to the three unsupervised clustering algorithms is the number of answers ("classes") that were proposed in the Galaxy Zoo DECaLS project. Even though some improvement could be brought to their method, they conclude that unsupervised classification can be useful to detect features and provide "human-like" classifications.

All of the unsupervised studies compare their results with some ad-hoc and simplified version of the de Vaucouleurs {stages} showing that this goal might be impossible to reach. Similarly to the supervised approaches, unsupervised classifications seem {to be more sensitive to} the features in the images than on the traditional morphological classification. {This is because the visual classifications have unavoidably a goal in mind (the physical significance) and thus get rid of a priori unrelated details.} This suggests the need for a new objective and automatic classification scheme. {But before embarking on such a big task, we should check whether it could }{yield} interesting insights into the physics of galaxies.

{This is exactly the goal of the present study: to }evaluate an unsupervised classification by interpreting the classes in a similar way as performed in \citet{deLapparent2011} on the EFIGI data, that is with the distribution of the physical properties of galaxies. It is not my goal here to propose a new general classification scheme based on images, but to show that a "traditional" statistical method compares favourably with unsupervised Deep Learning or visual approaches. The algorithm used here is a sub-space Gaussian mixture model called Fisher-EM  \citep{Bouveyron2012} that performs a dimensionality reduction optimized for the clustering. It has been used several times in astrophysics \citep{Siudek2018,Fraix-Burnet2021,Turner2021,Dubois2022}.

The paper is organized as follows. In Sect.~\ref{Data} the EFIGI data are presented together with the pre-processing performed for this study. In Sect.~\ref{section:method} the Fisher-EM algorithm is described. The results of the unsupervised classification are presented in Sect.~\ref{Results}. The comparison with Deep Learning studies and the physical interpretation of the Fisher-EM classification scheme are described in Sect.~\ref{Discussion}. In Sect.~\ref{Conclusion}, a conclusion from the present results on automatic classification in general is given with some perspectives in the frame of the data avalanche to come.

\section{Data}
\label{Data}

The EFIGI catalogue \citep{Baillard2011} is a multi-wavelength database specifically designed to densely sample all Hubble types. It includes 4458 galaxies, visually classified into 18 {types} by a group of  astronomers, and provides 16 attributes characterising the morphology of galaxies such as the bulge-to-total luminosity ratio and the colour {$(B-V)_T$ from the Principal Galaxy Catalogue \citep[hereafter PGC,][]{Paturel1995}}. The 18 {types of EFIGI} are a slightly modified version {of the 18 stages of the} Hubble classification by \citet{deVaucouleurs1959} {and used in the Third Reference Catalogue of Bright Galaxies \citep[RC3,][]{deVaucouleurs1991}.}

The EFIGI visual morphological classification was established {on "true" colour images made from the composite of images in the bands} $g, r$ and $i$, with a gamma correction and a "colour saturation exaggerated by a factor 2.0" optimised to help the human eyes to detect structures { in typical screen viewing conditions} \citep{Baillard2011}. These corrections are arbitrary, depend on the data set and the human classifiers, and are not necessarily suited for an algorithm.

{For our analyses, the $g, r, i$ images have been retrieved from the EFIGI website without any modification except the ones explained below. The central $100\times 100$ pixels part was considered to avoid foreground stars, companion galaxies or other artefacts as much as possible. This size is sufficient to encompass all galaxies without generating too much noisy background for the smallest ones. The images were rotated using Principal Component Analysis and hence orientated according to the axis of maximum variance, i.e. the dominant elongation. They were then centred using the brightest pixel in the very central region.}
The final images were finally transformed into vectors.

We have experimented with several alternatives, using the $r$ band only (plain or logarithmic),  normalised images, the concatenation of the vectors of the three bands, the concatenation of the $r$ band vector with the weighted vector for the colour $g-i${, and compare the results regarding their agreement with the reference classification from EFIGI.}

We have found that the logarithmic images provide a better segregation of the structures with our algorithm. The normalisation of the images has the drawback of increasing the noise for faint galaxies. Since the images are calibrated in mag arcsec$^2$, we decided not to normalise the images.

The analysis with the $r$ band alone is sensitive to the distribution of the light in the image, hence discriminates between face-on and edge-on galaxies. The concatenation of the three bands does not change the results.

The colour $g-i$ alone helps to identify irregular or very blue structures or galaxies, but is not able to well distinguish the inclination. As a consequence, the concatenation of the $r$ band with $g-i$ provides the best compromise. Weighting the $g-i$ image does not improve the result.

In the rest of this paper, we consider only the analysis with the logarithmic vectors of the concatenated $r$ band and $g-i$. It must be kept in mind that this choice is heavily related to the goal of the study, here to be close to the reference classification from EFIGI.

Our pre-processing of images is indeed minimalist, and could certainly be improved by reduction of noise, removal of foreground stars etc, but the long term objective is to analyse a very large sample of images to build a training sample and subsequently automatically classify a huge flow of data.

\section{The Fisher-EM algorithm}
    \label{section:method}

We have applied the unsupervised classification method called Fisher-EM \citep{Bouveyron2012} on the sample of images transformed into vectors. Fisher-EM is a subspace Gaussian mixture algorithm that relies on a statistical model, called the discriminative latent mixture (DLM) model. It uses a modified version of the expectation-maximisation (EM) algorithm by inserting a Fisher step to optimise the ratio of the sum of the between-class variance over the sum of the within-class variance for a better clustering.

Formally, we may define the observation vector-image $\bf{Y}=\{y_1, ..., y_n\}$ such that $y_i\in \nbR^{p}$ describes vector-image number $i$. The dimension $p$ corresponds to the $p$ fluxes at the $p$ pixels of the vector-image.

Classifying the observations into $K$ clusters mathematically translates into finding the vector $\bf{Z}=\{z_1, ..., z_n\}$, which assigns each vector-image $y_i$ to a given cluster $z_i \in  [\![1,K]\!]$. 
In the case of Fisher-EM, the clustering process occurs in a subspace $\nbE \subset \nbR^p$ of dimension $d=K-1 < p$. 
Therefore, the Gaussian mixture model is applied to the projected data $\bf{X}$ rather than the observed data $\bf{Y}$,
\begin{equation}
	\centering
	\bf{Y} = \bf{U}\bf{X} + \bf{\epsilon}
	\label{eq:Y}
	,\end{equation}
where $\bf{U} \in \mathcal{M}_{p,d}(\nbR)$ is the projection matrix and $\bf{\epsilon}$ is a noise vector of dimension $p$ following a Gaussian distribution centred around 0 and of covariance matrix $\Psi$ ($\varepsilon_{k}\sim\mathcal{N}(0,\Psi_{k})$). The multivariate Gaussian probability distribution $\bf{X}$ describing the cluster $k$ in the subspace is parametrised by a mean vector $\bf{\mu_k}$ and a covariance matrix $\bf{\Sigma}_k$,
\begin{equation}
	\centering
	X|_{Z=k} \sim \mathcal{N}(\mu_k, \Sigma_k)
	\label{eq:X}
	.\end{equation}
Combining Eqs.~\ref{eq:Y} and \ref{eq:X}, we obtain
\begin{equation}
	\centering
	Y_{|X,Z=k}\sim\mathcal{N}(UX,\Psi_{k})
	.\end{equation}
The observed data are thus modelled by a marginal distribution $f(\bf{y})$ that is the sum of $K$ multivariate Gaussian density functions $\phi$ of mean $\bf{U}\mu_k$ and covariance $\bf{U}\Sigma_k\bf{U^t} + \Psi$, each weighted by the corresponding mixing proportion $\pi_k$,
\begin{equation}
	\centering
	f(\bf{y}) = \sum_{k=1}^K \pi_k \phi(y;\bf{U}\mu_k, \bf{U}\Sigma_k\bf{U^t} + \Psi)
	\label{eq:f}
	.\end{equation}

By further assuming that the noise covariance matrix $\Psi_{k}$
satisfies the conditions $V^{t}\Psi_{k} V=\beta_{k}\mathbf{I}_{p-d}$ , where $V$ is the orthogonal complement of $U$, and $U^{t}\Psi_{k} U=\mathbf{0}_{d}$, the whole statistical model denoted by $\mathrm{DLM}_{[\Sigma_{k}\beta_{k}]}$
can be shown to take the following form: $$ \left(  \begin{array}{c@{}c} \begin{array}{|ccc|}\hline ~~ & ~~ & ~~ \\  & \Sigma_k &  \\  & & \\ \hline \end{array} & \mathbf{0}\\ \mathbf{0} &  \begin{array}{|cccc|}\hline \beta_{k} & & & 0\\ & \ddots & &\\  & & \ddots &\\ 0 & & & \beta_{k}\\ \hline \end{array} \end{array}\right)  \begin{array}{cc} \left.\begin{array}{c} \\\\\\\end{array} \right\}  & d \leq K-1\vspace{1.5ex}\\ \left.\begin{array}{c} \\\\\\\\\end{array}\right\}  & (p-d)\end{array}$$ These
last conditions imply that the discriminative and the non-discriminative
subspaces are orthogonal, which suggests in practice that all the
relevant clustering information remains in the latent subspace. From a practical point of view, $\beta_{k}$ models the variance of the non-discriminative noise of the data.

Several other models can be obtained from the DLM$_{[\Sigma_{k}\beta_{k}]}$ model by relaxing or adding constraints on model parameters. It can, for example, be assumed that the noise parameter $\beta_{k}$ differs from cluster to cluster, or that the covariance matrices $\Sigma_k$ are the same for all $K$ clusters. A thorough description of the DLM model, its 12 declinations, and the algorithm itself can be found in \cite{Bouveyron2012}.

The Fisher-EM algorithm sets the dimension of the latent subspace to be $d=K-1$. The number of clusters $K$ and the DLM model are given as input and the optimization loop iterates through the three following steps:
\begin{itemize}
	\item E-step: The posterior probabilities that the $n$ observations $y_i$ belong to each of the $K$ clusters are computed.
	\item Fisher-step: 
	The projection matrix $\bf{U}$ is computed to maximise the Fisher criterion.
	\item M-step: The DLM model parameters are adjusted to maximise the likelihood.
\end{itemize}

The choice of the best statistical DLM model and the optimum number of clusters $K$ depends on the data and was here estimated with the integrated completed likelihood (ICL) criterion. This criterion  penalises the likelihood by the number of parameters of the statistical model, the number of observations, and favours well-separated clusters \citep{Biernacki2000,Girard2016}. 

We used 25 random initialisations, and a sigmoid kernel to stabilize the algorithm since we are in the case $n < p$ ($n=4~458, p=10~000$). The Fisher-EM algorithm (version 1.6) is implemented in the eponym package for \textsf{R}.

\section{Results}
\label{Results}

\begin{figure}
	\includegraphics[width=\columnwidth]{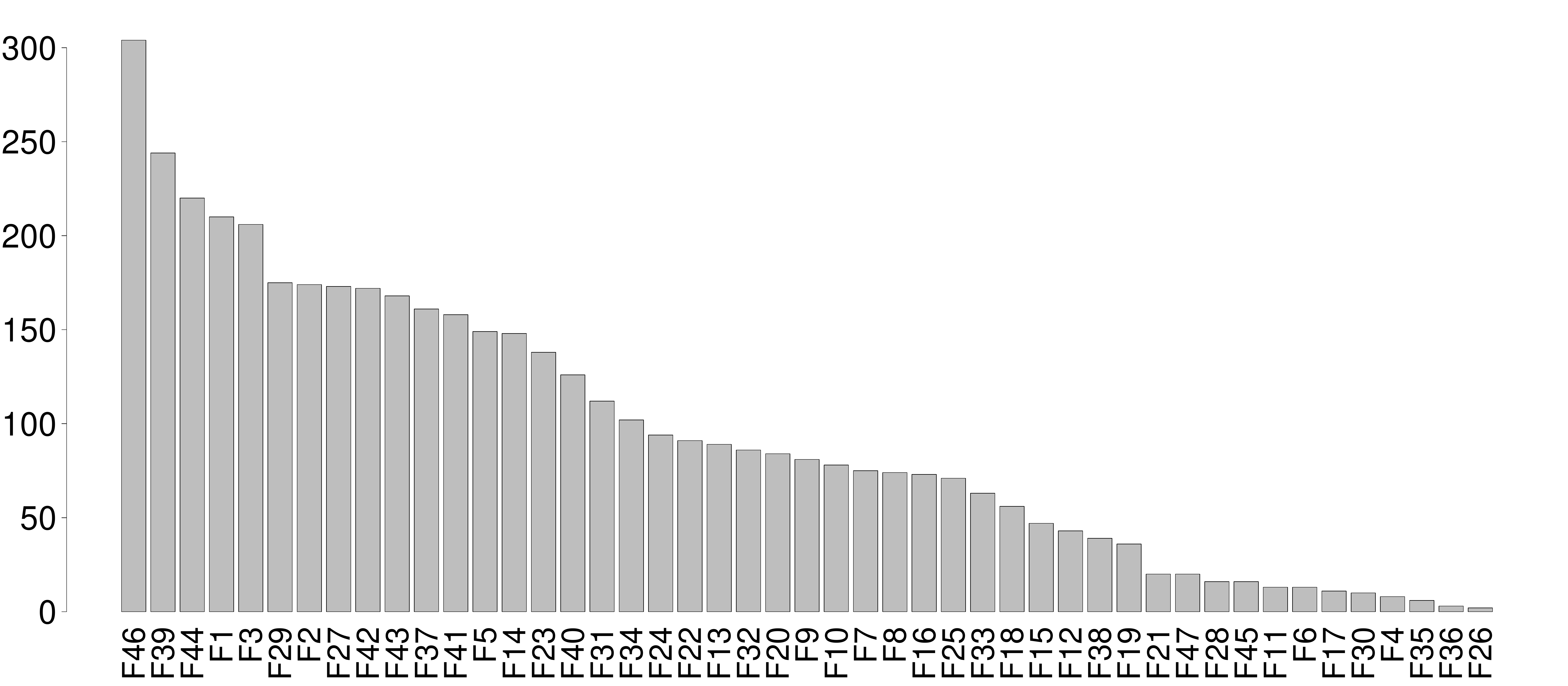}
		\caption{Ordered histogram of the number of galaxies per Fisher-EM class.}
	\label{fig:histogram}
\end{figure}

\begin{figure*}
	\includegraphics[width=\textwidth]{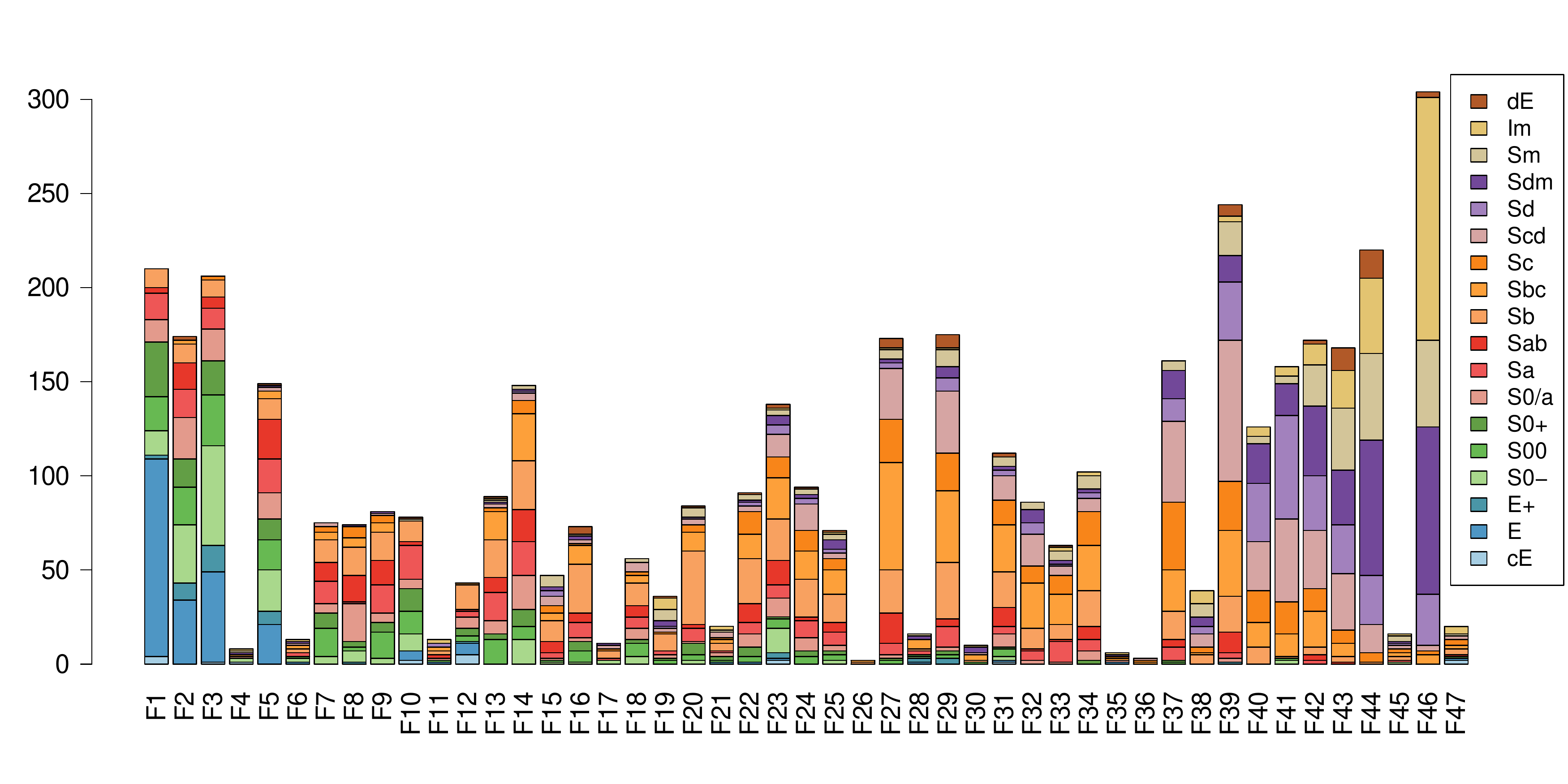}	\caption{Barplot showing the EFIGI {types} within each of the Fisher-EM class. The colour code of the former is given in the inlet at top right.}
	\label{fig:barplot}
\end{figure*}

\begin{figure}
	\includegraphics[width=\columnwidth]{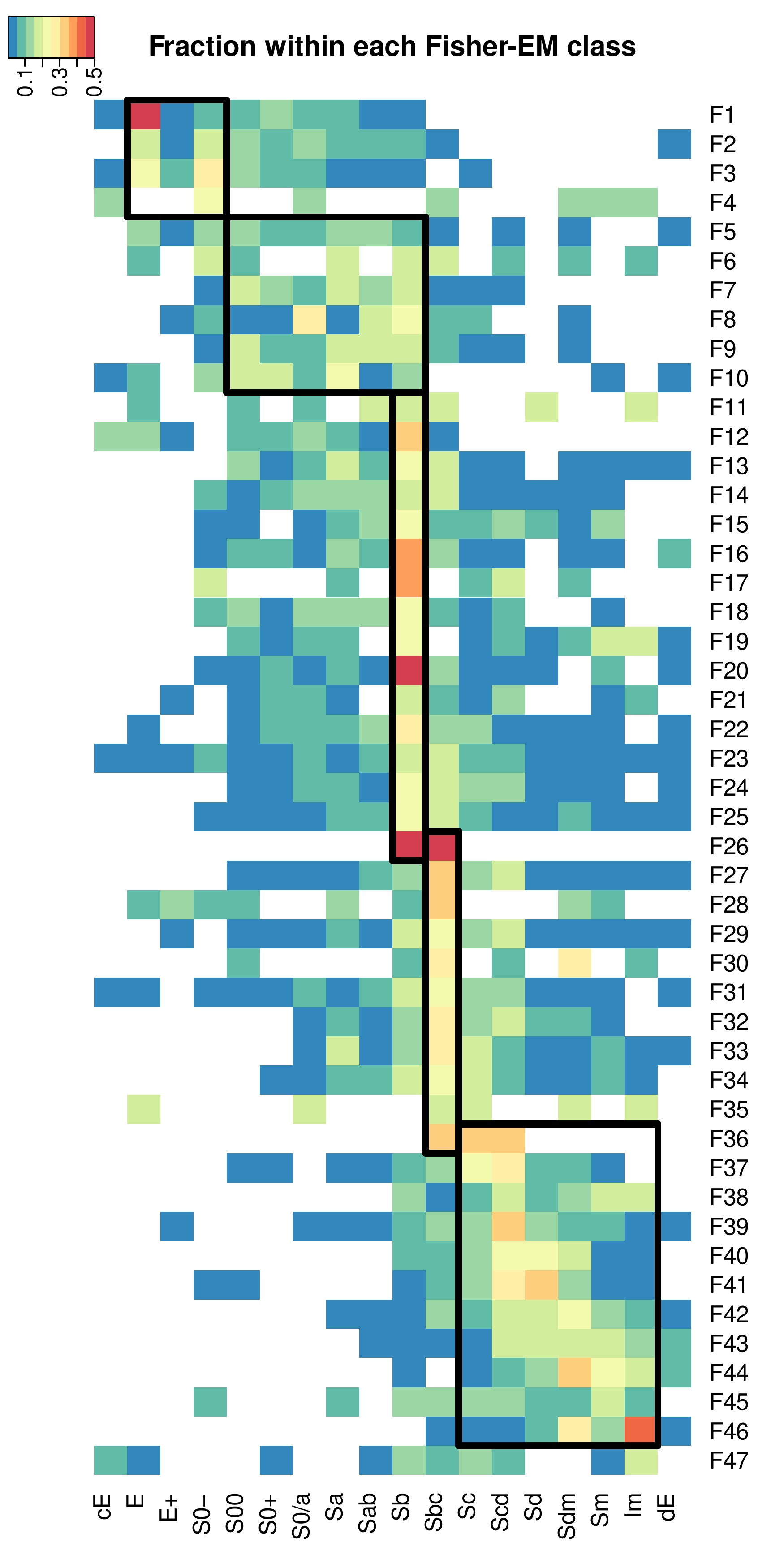}	
	\caption{Heatmap of the Fisher-EM classes versus the EFIGI {types}. The colour code given on top left corresponds to the fraction of the EFIGI {types} within each Fisher-EM class. In other words, the sum along the horizontal lines is 1, and this applies to all the heatmaps presented in this paper. The black boxes delineate five regions that are described in the text.}
	\label{fig:heatmapT}
\end{figure}

The optimum number of clusters is found to be 47, that is 2.6 times more than for the EFIGI classification. The number of elements per class (Fig.~\ref{fig:histogram}) goes from 2 (class F26) to 304 (class F46) with 12 classes having 20 or less images. 

The distribution of the EFIGI {types} in the Fisher-EM classes is depicted as a barplot in Fig.~\ref{fig:barplot}. Despite {being} far from a one to one correspondence between the Fisher-EM and the EFIGI {types}, the agreement is globally correct considering broad categories such as ellipticals/lenticulars, spirals or diffuse/irregulars as underlined by the colour coding. 
The contingency table (Table~\ref{tab:contingency_table}) gives a quantitative view of this correspondence and its dispersion. The heatmap (Fig.~\ref{fig:heatmapT}) showing the fraction of the EFIGI {types} for each Fisher-EM class better illustrates this result. 

Five zones can be identified in the contingency table and are indicated on the heatmap {with boxes drawn manually} (Fig.~\ref{fig:heatmapT}). The {types} E, E+ and S0- are gathered mainly in the classes F1 to F4. F1 is made of 50\% of EFIGI class E, but also 7\% of Sa and 5\% of Sb. Conversely about 50\% of the E are in F1 while the other ones are essentially in F2, F3 and F5.

The second zone is a diffuse distribution of lenticulars and Sa/Sb galaxies spread in classes F5 to F10.

The classes from F11 to F26 are dominated by Sb galaxies, while F26 to F36 are dominated by Sbc. 

Finally, the classes from F36 to F46 show a relatively fair agreement with the EFIGI {types} Sc to Im despite {the fact that} the corresponding galaxies have structures which are more disturbed or fainter. We note the very good concentration of Im galaxies in the class F46.

The conclusion at this level is that the correspondence between our Fisher-EM and the EFIGI classifications is relatively good if we consider only gross categories. The most remarkable result is that the Sb and Sbc {types} are divided into more than 10 Fisher-EM classes. To better understand the Fisher-EM classification, we propose a short visual description of the average properties of the images within the classes (Table~\ref{tab:properties}).

\begin{table*}
	\centering
	\caption{Visual short description of the Fisher-EM classes based on both the $r$ and png images. The horizontal lines delineate the five zones described in the text and shown in Fig.~\ref{fig:heatmapT}.}
	\label{tab:properties}
	\begin{tabular}{llll}
		\hline
Name & Number of  & Brief description & Approximate correspondence \\ 
     & members    &                   & with EFIGI classification \\
     &            &                   & (from Fig.~\ref{fig:heatmapT})\\
		\hline
F1 & 210 & 	Extended galaxy, bright bulge &	 E  \\
F2 & 174 & 	Slightly elongated &	 E S0-  \\
F3 & 206 & 	Bright core, less extended than the image &	 S0- E  \\
F4 & 8 & 	A very bright star (at bottom right) &	 S0-  \\
\hline
F5 & 149 & 	Bright core, extended, slightly elongated or inclined &	 E S0-  \\
F6 & 13 & 	A very bright star in the field (to the left) &	 S0- Sb Sbc  \\
F7 & 75 & 	Inclined, disky &	 S00 Sa Sb Sab  \\
F8 & 74 & 	Small bulge, face-on disk &	 S0/a  \\
F9 & 81 & 	Very inclined disk, small and bright bulge &	 Sa Sb S00 Sab  \\
F10 & 78 & 	Bright and large bulge, inclined or elongated &	 Sa  \\
\hline
F11 & 13 & 	Contaminated &	 ?  \\
F12 & 43 & 	Bright and large bulge &	 Sb  \\
F13 & 89 & 	Inclined disk, bright core &	 Sb  \\
F14 & 148 & 	Very inclined disk &	 Sb Sbc  \\
F15 & 47 & 	Edge-on weak spiral &	 Sb  \\
F16 & 73 & 	Large scale structure with bar or arms near the centre &	 Sb  \\
F17 & 11 & 	Bright star (bottom right) &	 Sb  \\
F18 & 56 & 	Nearly edge-on disk or bar &	 Sb  \\
F19 & 36 & 	Nearly coreless face-on disks &	 Sb  \\
F20 & 84 & 	Nearly face-on spiral, structured, extended, small core &	 Sb  \\
F21 & 20 & 	Inclined spiral, weak core, distorted &	 Sb  \\
F22 & 91 & 	Small face-on spirals, weak core &	 Sb  \\
F23 & 138 & Small disk, weak core &	 Sb Sbc  \\
F24 & 94 & 	Edge-on disk, weak bulge &	 Sb  \\
F25 & 71 & 	Face-on disk with bar or very inclined with weak bulge &	 Sb Sbc  \\
F26 & 2 & 	Spiral, a star at top right &	 ?  \\
\hline
F27 & 173 & 	Face-on weak spiral, weak bulge &	 Sbc  \\
F28 & 16 & 	A bright star in the field (to the left of the image) &	 Sbc  \\
F29 & 175 & 	Face-on weak spiral with strong arms &	 Sbc  \\
F30 & 10 & 	Bright star at top left. &	 Sbc Sdm  \\
F31 & 112 & 	Face-on weak spiral with small core &	 Sbc  \\
F32 & 86 & 	Edge-on galaxy with small bulge &	 Sbc  \\
F33 & 63 & 	Very slightly inclined spiral or elongated galaxy, weak core  &	 Sbc  \\
F34 & 102 & 	Very inclined weak spiral, weak core &	 Sbc Sb Sc  \\
F35 & 6 & 	Small star close to the centre &	 ?  \\
F36 & 3 & 	Weak face-on spiral, weak core, with a star in the field &	 Sbc Sc Scd  \\
\hline
F37 & 161 & 	Edge-on galaxies, very weak core &	 Scd  \\
F38 & 39 & 	Nearly edge-on weak disk, no core &	 Scd Sdm Im  \\
F39 & 244 & 	Nearly coreless face-on faint spiral with conspicuous arms &	 Scd  \\
F40 & 126 & 	Edge-on weak disk, very small bulge &	 Sd  \\
F41 & 158 & 	Edge-on weak disk, no bulge &	 Sd  \\
F42 & 172 & 	Very inclined weak disk &	 Sdm  \\
F43 & 168 & 	Very faint inclined disk &	 Sm Scd Sdm Sd  \\
F44 & 220 & 	Very faint face-on patchy disks &	 Sdm  \\
F45 & 16 & 	Contamination by star or another galaxy &	 Sm  \\
F46 & 304 & 	Faint, irregular or diffuse &	 Im  \\
F47 & 20 & 	Complex structure &	 Im Sb Sc  \\
		\hline
\end{tabular}
\end{table*}

\begin{figure*}
	\includegraphics[width=\textwidth]{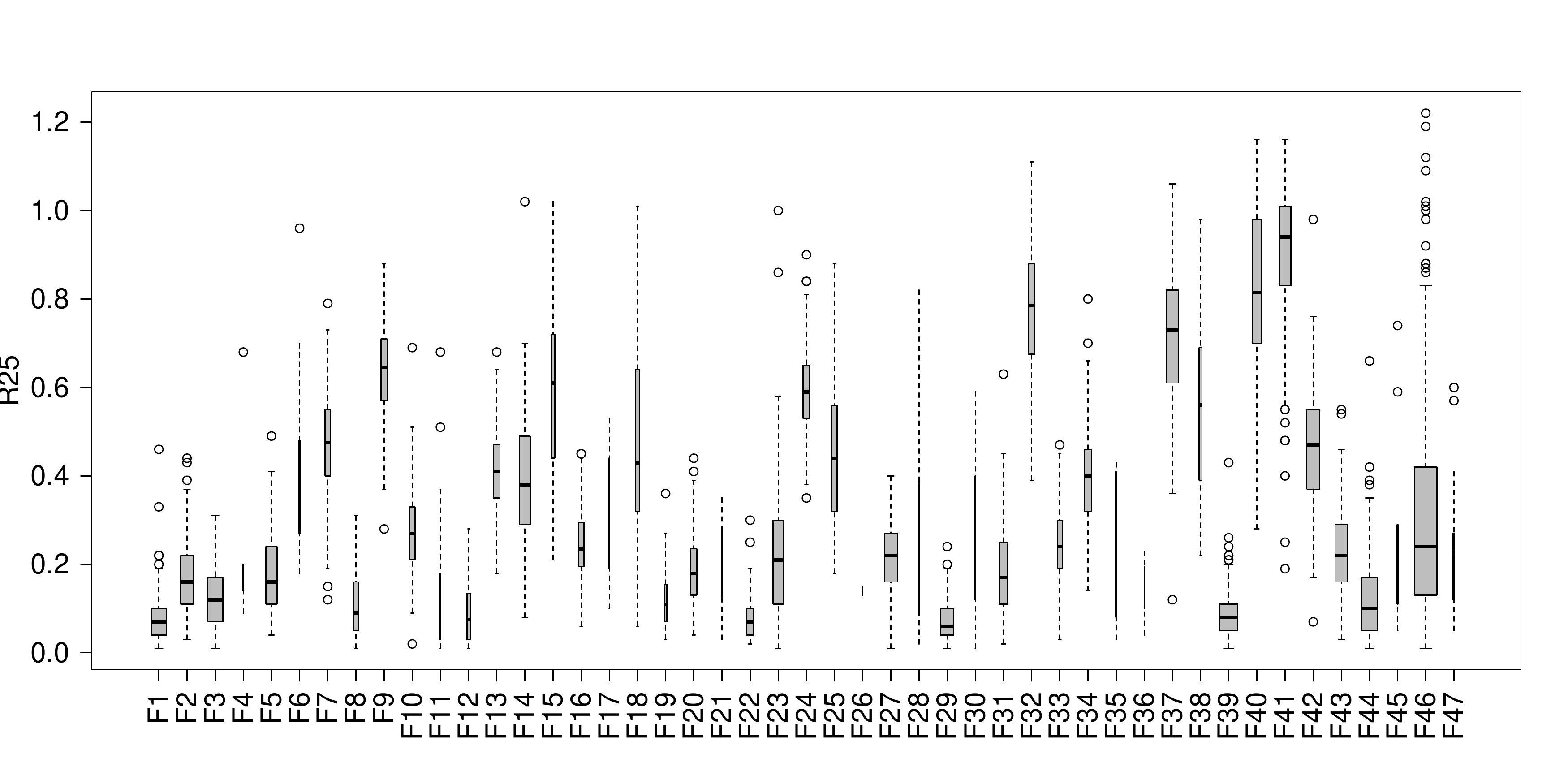}
		\caption{Boxplot of the parameter R25 \citep{Baillard2011} characterizing the ratio between major and minor isophotal diameters (see text). The height of the boxes show the quartiles, and their width is proportional to the size of the class.}
	\label{fig:R25}
\end{figure*}

\begin{figure*}
	\includegraphics[width=\columnwidth]{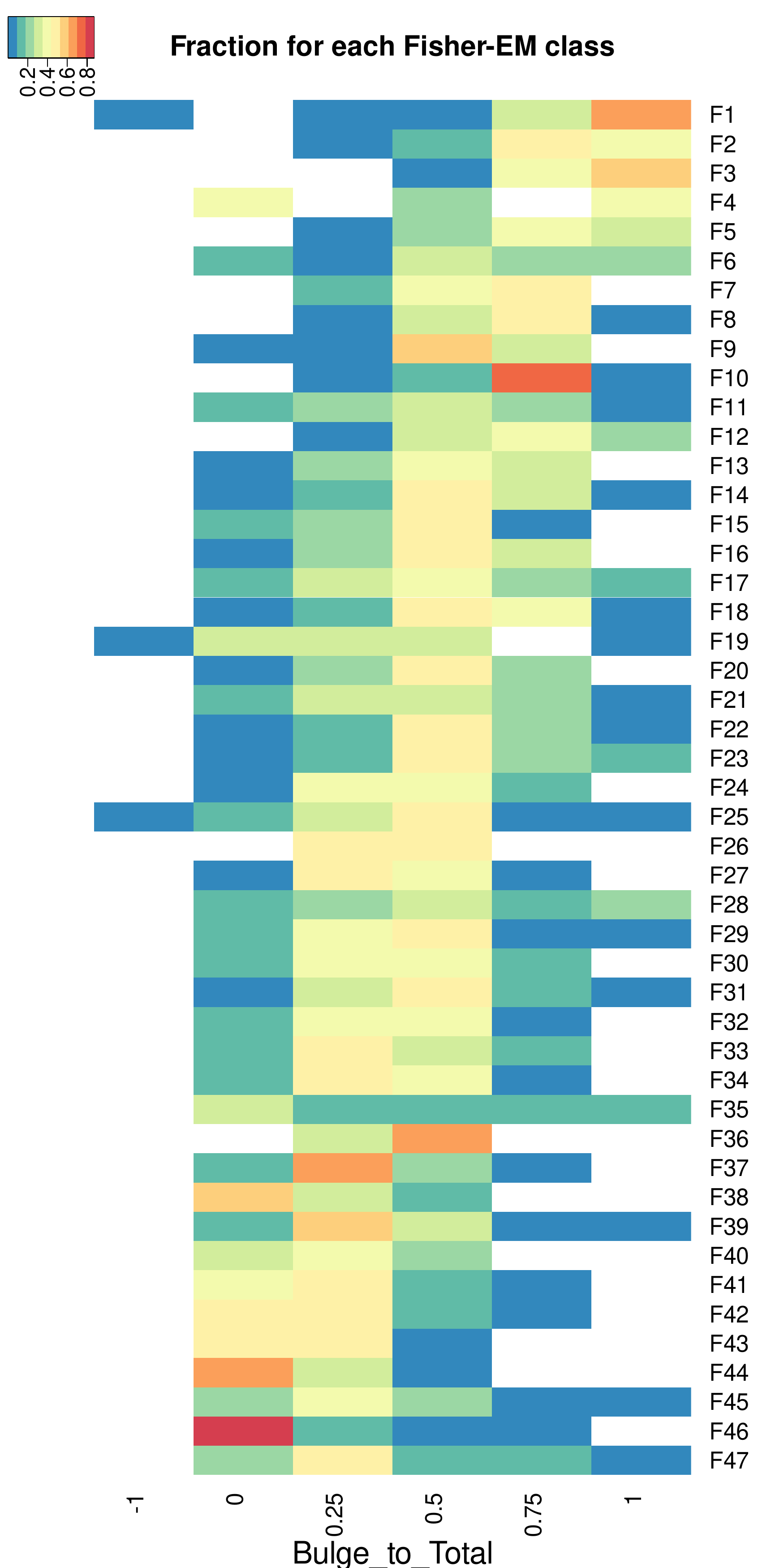}
\includegraphics[width=\columnwidth]{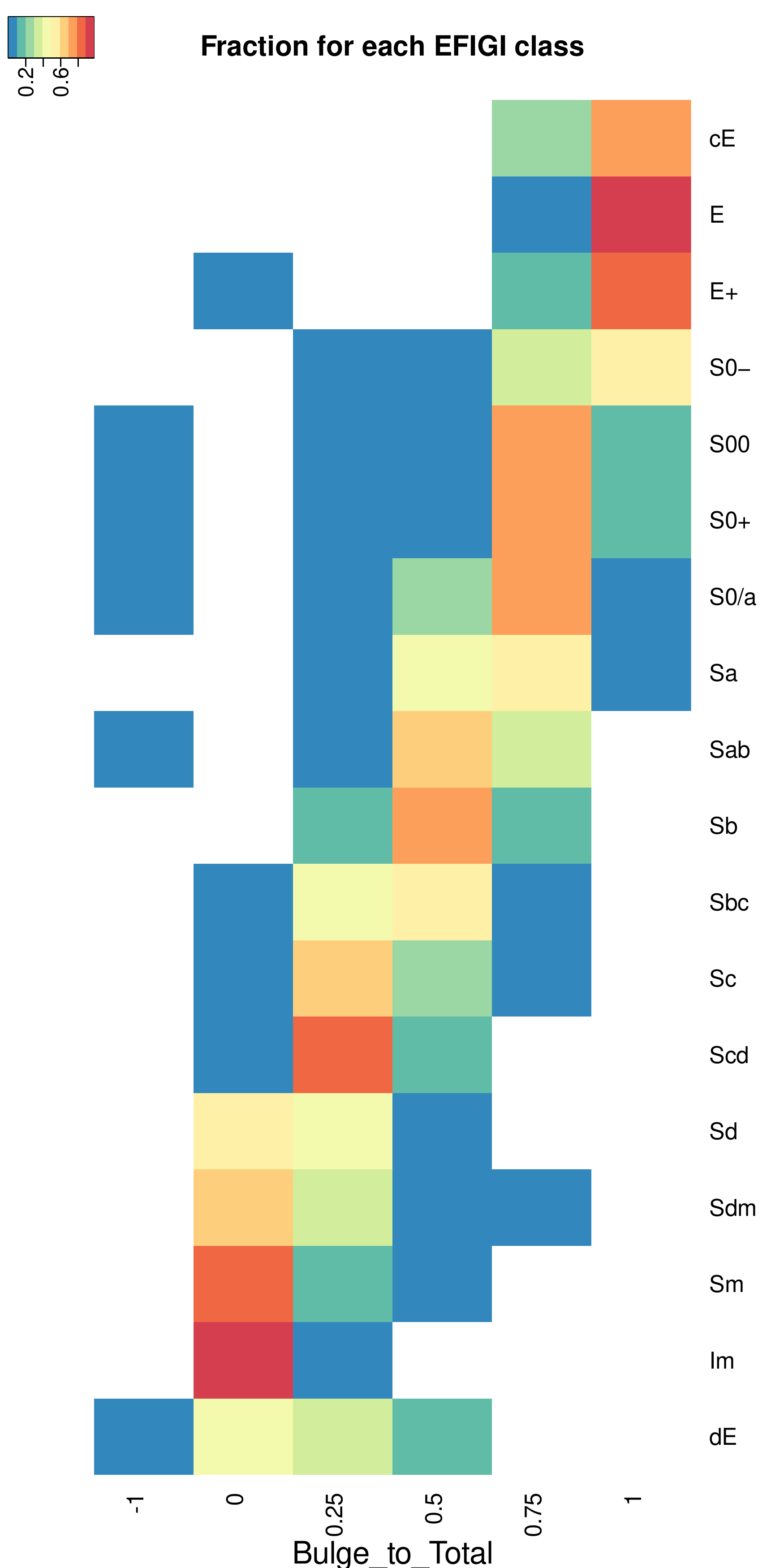}
	\caption{Heatmap of the bulge to total ratio for the Fisher-EM classification (left) and the EFIGI classification (right). Note that the colour codes are not exactly the same (see inlets).}
	\label{fig:heatmapBtT}
\end{figure*}

Globally, we find that the distribution of the light in the images is the main discriminative characteristics explaining the Fisher-EM classification. The inclination or elongation is distinguished between face-on spirals (e.g. F8, F19...), inclined or elongated galaxies (F7, F9, F13) or edge-on disks (F14, F15, F18).  There are even different levels of inclination (F13, F14, F15). This is confirmed by the PGC parameter R25 (Fig.~\ref{fig:R25}) which is the decimal logarithm of the ratio between the mean major isophotal diameter and the mean minor isophotal diameter measured at or reduced to the surface brightness level 25.0 mag.arcsec$^{-2}$ in the B band \citep{Baillard2011}. Each Fisher-EM class shows relatively little dispersion on this parameter indicating a very good discrimination. 

The prominence of the bulge is well discriminated independently of the galaxy inclination (Fig.~\ref{fig:heatmapBtT}). Contamination is detected (F11), and remarkably enough, the presence of a star in the field is detected and discriminated according to its position: top-left (F30), top-right (F26), bottom-right (F4, F17), left (F6, F28). These classes with a star all have less than 20 members.

The detailed structures are also well retrieved and discriminated. The class F46, the largest with 304 members, gathers most of the galaxies classified as irregulars (Im) as well as many Sdm and Sm galaxies described by \citet{Baillard2011} as <<very loosely wound or with some indication of spiral arms, very weak bulge, low amounts of dust>>.

Some of our classes are made exclusively of contaminated images. This is confirmed by the corresponding EFIGI attributes. The Contamination attribute goes from 0 to 1 with increasing importance of the contaminant \citep{Baillard2011}. Six of the Fisher-EM classes (F6, F26, F28, F30, F35, F36) have non-zero Contamination values (F6, F26 and F30 having even values higher than 0.5) while all EFIGI {types} have values from 0 to 0.75. For the attribute Multiplicity, 12 of our classes have  values 0 (no other galaxy) and 0.25 (one neighbouring galaxy), one class (F36) has 0 only, while for the EFIGI {types} this attribute has always at least values of 0, 0.25 and 0.5 (two neighbouring galaxies).

As a conclusion, the high number of classes found by Fisher-EM is due to the detailed discrimination in the light distribution and in the presence of specific features within the images.

\begin{figure}
	\includegraphics[width=\columnwidth]{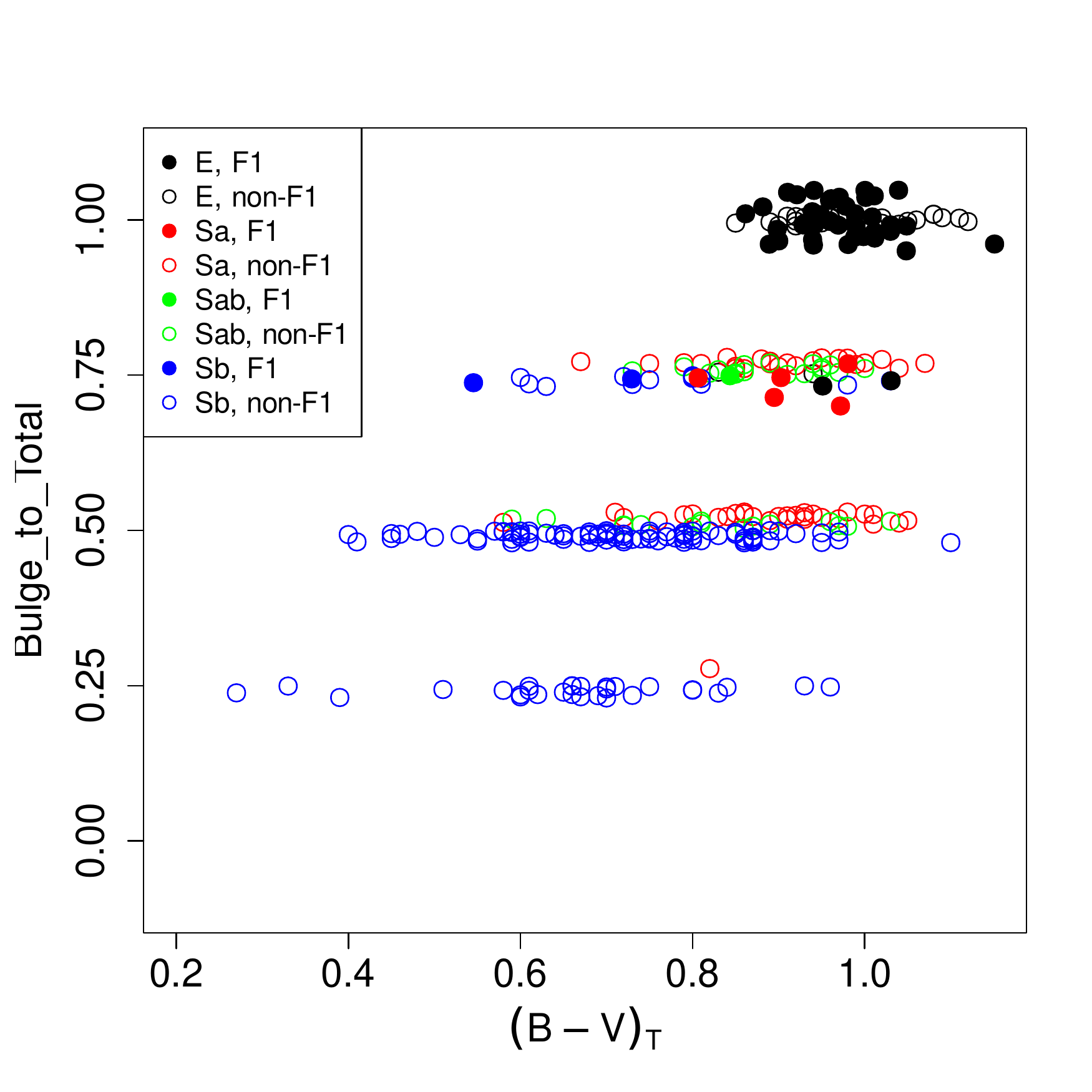}	\caption{Bulge to total luminosity ratio vs $(B-V)_T$ for the ellipticals E (black points), Sa (red), Sab (green) and Sb (blue) galaxies in the EFIGI sample. Filled circles are for the members of the F1 class while the open circles are for all the others. A jiggle has been added to separate the points.}
	\label{fig:F1Spirals}
\end{figure}

In this context, the presence of spirals Sa, Sab and Sb in the class F1 which is dominated by ellipticals (50\%) could seem contradictory. There are only 3 Sab galaxies but Sa galaxies represent 7\% of the class F1 and the Sb ones 5\%. This is not negligible and the most immediate conclusion would be that the algorithm is not able to detect relatively faint spiral features. However, these galaxies have a high bulge to total luminosity ratio and a low $(B-V)_T$ as compared to the other Sa, Sab and Sb spirals in this sample, making them somewhat closer to ellipticals  (Fig.~\ref{fig:F1Spirals}). In addition, they have a marginally lower arm strength than the other spirals in the EFIGI sample (Fig.~\ref{fig:heatmapArmStrength}).

\section{Discussion}
\label{Discussion}

\subsection{Comparison with other results}
\label{comp}

Our Fisher-EM classification shows a reasonable agreement with the EFIGI classification (Fig.~\ref{fig:heatmapT}). To our knowledge, our work is the first to provide a comparison with the full de Vaucouleurs 18 {stages}. Despite a very simple pre-processing of the images, our state-of-the-art "traditional" Machine Learning clustering tool appears to perform similarly to more complicated Deep Learning techniques that consider a simplified classification scheme grossly corresponding to our five zones identified as black boxes in Fig.~\ref{fig:heatmapT} (see Sect.\ref{Introduction}).

In all cases, it must be kept in mind that the EFIGI classification has been obtained on png composite colour images made with the $r, g$ and $i$ bands with a gamma correction and a saturation optimized for the rendering of the objects, that is for the human eye \citep{Baillard2011}. There is also an intrinsic uncertainty in the EFIGI classification (Fig.~\ref{fig:heatmapTunc}), and above all, there is an intrinsic fuzziness in the class definitions (Sect.~\ref{Introduction}).
As a consequence, it cannot be expected for any automatic tool to compete with these very human judgements and perfectly match the visual morphological classification. All the unsupervised and supervised Machine Learning classification studies have reached this same conclusion, ours included. 

Our detailed classification confirms that the algorithms in general are more sensitive to the distribution light in the images than the de Vaucouleurs classification which is based principally on the properties of the spiral arms and the presence of a bulge, the human brain "deconvolving" characteristics such as disk inclination or foreground stars. Even if not stated as clearly, this finding is present in all citizen and automatic classification studies  (Sect.~\ref{Introduction}).

Our Fisher-EM analysis also confirms that the actual number of classes of galaxy images is significantly higher than 18. We find an optimum of 47 classes using an objective criterion (ICL), which corresponds to a latent subspace dimension of 46. This is to be compared with the 27 clusters found with Variational Auto-Encoder by \citet{Cheng2021} and the 37 categories of the citizen science project Galaxy Zoo 2 \citep{Willett2013}. Unfortunately, only 13 galaxies are common between the EFIGI and the Galaxy Zoo 2 samples. The comparison is difficult since the latter does not provide a classification but a description of the galaxies. Still, we find a very good correspondence.



\subsection{Physical interpretation of the Fisher-EM classification}

The present study is not intended to provide a new classification scheme of the morphology of galaxies. For that, one should analyse a much larger sample {to increase the diversity of features. However it should be useful to clarify the goal}: building feature-based catalogues or gaining physical insights? In this section, we show that both are compatible: unsupervised classification can provide relevant insights into the physics of galaxies equivalently to those provided through a careful visual classification. In this purpose, we follow the analysis of the EFIGI data by \citet{deLapparent2011}.

In our Fisher-EM classification, the inclination or elongation is very well discriminated (Fig.~\ref{fig:R25}), somewhat better than for the EFIGI classification since the dispersion per class is slightly smaller (Fig.~\ref{fig:heatmapInclination}). This parameter is quite important, for instance in the study of detailed structures of bulges which are easier for edge-on galaxies \citep{Laurikainen2018}.

The bulge to total luminosity ratio compares very well between our Fisher-EM and the EFIGI classifications (Fig.~\ref{fig:heatmapBtT}). Despite a slightly larger dispersion for the Fisher-EM classification, the main trend is very similar in both cases, meaning that the relative importance of the bulge is recovered. \citet{deLapparent2011} comment that the spread for the bulge-to-total ratio is due to the fact that "the Hubble sequence is primarily based on the strength and pitch angle of the spiral arms, independently from the bulge-to-total ratio". However, this is true for spiral galaxies only since the relative size of the bulge obviously decreases by construction from the ellipticals to the spirals. The lenticulars are of an intermediate nature the origin of which is still not understood \citep{Cavanagh2023}. The dispersion in the bulge to total luminosity ratio in the Fisher-EM classification is probably due to its multivariate nature that does not give any a priori weight to any feature in the images. In addition, many spirals have a very bright bulge with respect to the disk (Fig.~\ref{fig:F1Spirals}). In any case, the bulge-to-disk ratio is a widely used morphological quantitative index \citep[e.g.][]{Papaderos2022,Jang2023} and is an essential ingredient of the formation and evolution of galaxies, probably more than spiral arms which mainly indicates the presence of molecular gas and star formation \citep{Querejeta2021}. Our Fisher-EM classification is relatively well discriminative for this parameter.

The presence of a disk within ellipticals greatly complicates the simple picture of the classical morphological classifications. \citet{Graham2019} presents an extensive historical review of different attempts to build a synthetical diagram of the diversity of galaxy structures. Somehow abandoning the goal that such a diagram could depict the evolutionary pathways of galaxies, \citet{Graham2019} proposes that a grid could better fit the reality. This complexity probably explains the dispersion of the bulge-to-total luminosity ratio in our Fisher-EM classification. Interestingly, he notes that one benefit of a galaxy classification scheme is that "it creates familiarity with the morphology, structure, and components that galaxies are composed of", in agreement with \citet{Buta2019}. We can add that the algorithms, unsupervised or supervised, are very sensitive to features on the images as well (Sect.~\ref{comp}), and provide a different and unbiased view of the morphologies of galaxies.

Although the Hubble, de Vaucouleurs or EFIGI classifications seem primarily based on the properties of the spiral arms such as their strength and curvature \citep{deLapparent2011}, the spread in the arm strength is relatively large for the EFIGI classification, confirming that the spiral arms are indeed not the only criterion for the de Vaucouleurs classification. This spread is very comparable in the Fisher-EM classification (Fig.~\ref{fig:heatmapArmStrength}), implying that there is no real advantage to make a visual classification for this conspicuous feature with respect to an automatic unsupervised classification.

The colour $(B-V)_T$ is not explicitly taken into account in the EFIGI classification but it has been indirectly integrated through the colour images used and optimised for the human eye \citep{Baillard2011}. We know that this property grossly correlates with the level of star formation and dust, hence with the de Vaucouleurs sequence through the relative importance between the bulge and the disk \citep{Querejeta2021}. In the Fisher-EM classification, the colour $g-i$ was included in the analysis in order to improve the distinction between ellipticals and spirals. Unsurprisingly, the same trend is found for $(B-V)_T$ along both the Fisher-EM sequence and the EFIGI one (Fig.~\ref{fig:heatmapBmV}).

The intrinsic size of the galaxy is not explicitly present in the EFIGI classification but it is known to be specific to some types of galaxies. We do not repeat here the detailed discussion by \citet{deLapparent2011} but we simply note that the diameter D25 (in kpc) follows exactly the same trend in our Fisher-EM classification  (Fig.~\ref{fig:heatmapD25kpc}).

The flocculence is a good indicator of the sensitivity of an algorithm to detect somewhat subtle structures within the galaxies. This indicator is indirectly included in the EFIGI classification and is naturally associated to star formation regions or perturbations. Flocculence is slightly more dispersed for our Fisher-EM classification, but the overall trend is very similar (Fig.~\ref{fig:heatmapFlocculence}). The same result is obtained for the presence of visible dust (Fig.~\ref{fig:heatmapVisibleDust}).

The conclusion of this section is that our unsupervised Fisher-EM classification yields very similar physical insights to those of the visual traditional morphological classification. It even appears more detailed since it has more classes corresponding to the distribution of light on the images.

\section{Conclusion}
\label{Conclusion}

Automatic classification of galaxies is recognised to be absolutely necessary to catalogue the huge number of new data that will soon feed astronomical data bases. Many developments have been performed in the last two decades, and in this paper I propose to draw some conclusions from a few representative results found in the literature, and from a new unsupervised analysis of the EFIGI galaxy images.

The general motivation of the first studies was to reproduce the visual classification, mainly the original Hubble scheme in four classes. Because it proved very difficult to match more detailed classifications such as the de Vaucouleurs 18-class scheme, different studies moved away from these very subjective {ways} of describing galaxies and concentrated more of the detection of features in the images. The most significant  move was the citizen science project Galaxy Zoo that abandoned the Hubble scheme between the first and second versions although both are entirely based on visual inspection. This is especially important because this project could have provided a training sample for supervised classification such as Deep Learning approaches. And the latter have basically found that using visual training samples does not provide {results as good as hoped for}. 

This may be why unsupervised clustering has gained some more interest in the very last years. This is also probably triggered by the occurrence of unsupervised Deep Learning and the Variational Auto-Encoders that allow dimensionality reduction without a training sample, opening the possibility of classification in this new data representation.

In this paper, I used a state-of-the-art "traditional" unsupervised clustering tool, Fisher-EM, based on Gaussian Mixture Models in a latent subspace, to analyse a sample of galaxy images which has been carefully examined and classified by expert astronomers in the EFIGI project following the de Vaucouleurs scheme.

The comparison between the 47 classes found by Fisher-EM and the 18 EFIGI {types} is reasonably good. This may seem somewhat disappointing but as recognised by many authors, one does not expect a perfect match between algorithms' results and the de Vaucouleurs classification since the latter is very subjective with definitions of categories being relative. The high number of classes found by Fisher-EM is in agreement with other results, in particular with the 37 categories of the Galaxy Zoo 2 project and the unsupervised Deep Learning study by \citet{Cheng2021} that finds 27 classes.

In this paper, I show that the algorithms are more sensitive to the distribution of light in the images, such as inclination, bulge to disk ratio, the presence of foreground stars or other features. This explains the higher number of classes with respect to visual classifications.

The most important result is that the unsupervised classification yields very similar insights into the physics of galaxies as compared to the visual de Vaucouleurs classification. The distribution of physical parameters such as $(B-V)_T$, bulge-to-total luminosity ratio, size, or even flocculence and dust, is similar in the Fisher-EM classes and in the EFIGI {types}.

In these conditions, should we really continue trying to reproduce precisely with an algorithm what the human eyes are able to do knowing that humans do not always agree on a classification for a given galaxy? Should we consider that the Machine Learning algorithms see the data differently but possibly in a more reproducible way? And couldn't this different vision of the data reveal something we could not imagine?

This raises the question of the kind of information that should be included automatically into the catalogues, and for what purpose. As shown in this paper, the algorithms distinguish patterns that are not necessarily useful to understand the evolution of galaxies, such as inclination or foreground stars, but may be useful for some other physical purposes (better characterisation of bulges in edge-on galaxies) or observational or technical information. 

And if the physical insight from an unsupervised Machine Learning classification is informative enough as shown in the present paper, then it is {probably time} to devise a new classification of galaxies based on these techniques that do not limit to Deep Learning and Gaussian Mixture Models mentioned in this paper \citep{Fraix-Burnet2015}. Each technique may be more or less suited depending on the data. The incredible images from the JWST make us disconcerted with regard to the description of structures and certainly appeal for new directions of classifying morphologies of galaxies. 

It is clear that galaxies are not described solely by images and morphologies in the visible domain, there {is so much more} information available. Depending on the goal of the classification, different variables/observables can be used. The evolution of galaxies does not depend on the number of spiral arms, but rather on the presence of a bulge, a disk, a perturbation, a stellar formation and a nuclear activity. Many of these properties can be determined automatically from the images \citep[e.g.][]{Ghosh2022} or from other instruments \citep[e.g.][]{Fraix2012,Chattopadhyay2019,Fraix-Burnet2021}.

In {conclusion}, while for a century we have tried to associate \emph{qualitative} categories of galaxy morphologies with \emph{quantitative} properties (bulge-to-disk ratios, colours, stellar formation...), it may be a good idea to do the reverse, i.e. characterise \emph{quantitative} categories obtained through multivariate unsupervised tools with \emph{qualitative} morphological description as illustrated in Table~\ref{tab:properties}? It could help to describe the evolution and formation of galaxies from multivariate physical properties \citep{Fraix2012} without putting too much emphasis on a specific feature, how interesting and attractive it may be.



\section*{Data Availability}

The data are available on the EFIGI website\footnote{\url{https://www.astromatic.net/projects/efigi/}}. The Fisher-EM algorithm is available as a package in the R environment\footnote{\url{https://cran.r-project.org/web/packages/FisherEM/index.html}}. Three files associated with this paper are available on \url{https://cloud.univ-grenoble-alpes.fr/s/EjW2fb6tMtxn5jC}: a table giving the Fisher-EM class for the 4458 galaxies and two pdf files with the png and the $r$ images for each Fisher-EM class.



\bibliographystyle{mnras}
  \bibliography{FEMmorpho} 
  
\appendix

%
%
%

\section{Additional table and figures for the EFIGI sample}

\begin{table*}
	\centering
	\caption{Contingency table between the Fisher-EM classification and the EFIGI classification. Numbers in boldface indicates the maximum over each row (Fisher-EM class).}
	\label{tab:contingency_table}
	\begin{tabular}{rrrrrrrrrrrrrrrrrrrr}
		\hline
		& cE & E & E+ & S0- & S00 & S0+ & S0/a & Sa & Sab & Sb & Sbc & Sc & Scd & Sd & Sdm & Sm & Im & dE & \textit{Total} \\ 
		\hline
		F1 & 4 & \textbf{105} & 2 & 13 & 18 & 29 & 12 & 14 & 3 & 10 &  &  &  &  &  &  &  &   & \textit{210} \\ 
		F2 &  & \textbf{34} & 9 & 31 & 20 & 15 & 22 & 15 & 14 & 10 & 2 &  &  &  &  &  &  & 2  & \textit{174} \\ 
		F3 & 1 & 48 & 14 & \textbf{53} & 27 & 18 & 17 & 11 & 6 & 9 &  & 2 &  &  &  &  &  &   &  \textit{206}\\ 
		F4 & 1 &  &  & 2 &  &  & 1 &  &  &  & 1 &  &  &  & 1 & 1 & 1 &   & \textit{8} \\ 
		F5 &  & 21 & 7 & \textbf{22} & 16 & 11 & 14 & 18 & 21 & 11 & 4 &  & 2 &  & 1 &  &  & 1  & \textit{149} \\ 
		F6 &  & 1 &  & 2 & 1 &  &  & 2 &  & 2 & 2 &  & 1 &  & 1 &  & 1 &   & \textit{13} \\ 
		F7 &  &  &  & 4 & \textbf{15} & 8 & 5 & 12 & 10 & 12 & 4 & 3 & 2 &  &  &  &  &   & \textit{75} \\ 
		F8 &  &  & 1 & 6 & 2 & 3 & \textbf{20} & 1 & 14 & 15 & 5 & 6 &  &  & 1 &  &  &   & \textit{74} \\ 
		F9 &  &  &  & 3 & 14 & 5 & 5 & \textbf{15} & 13 & \textbf{15} & 5 & 4 & 1 &  & 1 &  &  &   & \textit{81} \\ 
		F10 & 2 & 5 &  & 9 & 12 & 12 & 5 & \textbf{18} & 2 & 11 &  &  &  &  &  & 1 &  & 1  & \textit{78} \\ 
		F11 &  & 1 &  &  & 1 &  & 1 &  & 2 & 2 & 2 &  &  & 2 &  &  & 2 &   & \textit{13} \\ 
		F12 & 5 & 6 & 1 &  & 3 & 4 & 6 & 3 & 1 & \textbf{13} & 1 &  &  &  &  &  &  &   & \textit{43} \\ 
		F13 &  &  &  &  & 13 & 3 & 7 & 15 & 8 & \textbf{20} & 15 & 2 & 2 &  & 1 & 1 & 1 & 1  & \textit{89} \\ 
		F14 &  &  &  & 13 & 7 & 9 & 18 & 18 & 17 & \textbf{26} & 25 & 7 & 4 & 1 & 1 & 2 &  &   & \textit{148} \\ 
		F15 &  &  &  & 1 & 1 &  & 1 & 3 & 6 & \textbf{11} & 4 & 4 & 5 & 3 & 2 & 6 &  &   & \textit{47} \\ 
		F16 &  &  &  & 1 & 6 & 5 & 2 & 8 & 5 & \textbf{26} & 10 & 1 & 2 &  & 2 & 1 &  & 4  & \textit{73} \\ 
		F17 &  &  &  & 2 &  &  &  & 1 &  & \textbf{4} &  & 1 & 2 &  & 1 &  &  &   & \textit{11} \\ 
		F18 &  &  &  & 4 & 7 & 2 & 6 & 6 & 6 & \textbf{12} & 4 & 2 & 5 &  &  & 2 &  &   & \textit{56} \\ 
		F19 &  &  &  &  & 2 & 1 & 2 & 2 &  & \textbf{9} &  & 1 & 2 & 1 & 3 & 6 & 6 & 1  & \textit{36} \\ 
		F20 &  &  &  & 2 & 3 & 6 & 1 & 7 & 2 & \textbf{39} & 10 & 4 & 3 & 1 &  & 5 &  & 1  & \textit{84} \\ 
		F21 &  &  & 1 &  & 1 & 2 & 2 & 1 &  & \textbf{4} & 2 & 1 & 3 &  &  & 1 & 2 &   & \textit{20} \\ 
		F22 &  & 1 &  &  & 3 & 5 & 7 & 6 & 10 & \textbf{24} & 13 & 12 & 3 & 2 & 1 & 3 &  & 1  & \textit{91} \\ 
		F23 & 2 & 1 & 3 & 13 & 5 & 1 & 10 & 7 & 13 & \textbf{22} & \textbf{22} & 11 & 12 & 5 & 5 & 3 & 1 & 2  & \textit{138} \\ 
		F24 &  &  &  &  & 4 & 3 & 7 & 9 & 2 & \textbf{20} & 15 & 11 & 14 & 3 & 2 & 3 &  & 1  & \textit{94} \\ 
		F25 &  &  &  & 2 & 3 & 2 & 3 & 7 & 5 & \textbf{15} & 13 & 6 & 3 & 2 & 5 & 3 & 1 & 1  & \textit{71} \\ 
		F26 &  &  &  &  &  &  &  &  &  & 1 & 1 &  &  &  &  &  &  &   & \textit{2} \\ 
		F27 &  &  &  &  & 2 & 1 & 2 & 6 & 16 & 23 & \textbf{57} & 23 & 27 & 3 & 2 & 5 & 1 & 5  & \textit{173} \\ 
		F28 &  & 1 & 2 & 1 & 1 &  &  & 2 &  & 1 & \textbf{5} &  &  &  & 2 & 1 &  &   & \textit{16} \\ 
		F29 &  &  & 3 &  & 2 & 2 & 2 & 11 & 4 & 30 & \textbf{38} & 20 & 33 & 7 & 6 & 9 & 1 & 7  & \textit{175} \\ 
		F30 &  &  &  &  & 1 &  &  &  &  & 1 & \textbf{3} &  & 1 &  & \textbf{3} &  & 1 &   & \textit{10} \\ 
		F31 & 1 & 1 &  & 2 & 4 & 1 & 7 & 4 & 10 & 19 & \textbf{25} & 13 & 13 & 3 & 2 & 5 &  & 2  & \textit{112} \\ 
		F32 &  &  &  &  &  &  & 2 & 5 & 1 & 11 & \textbf{24} & 9 & 17 & 6 & 7 & 4 &  &   & \textit{86} \\ 
		F33 &  &  &  &  &  &  & 1 & 11 & 1 & 8 & \textbf{16} & 10 & 5 & 1 & 2 & 5 & 2 & 1  & \textit{63} \\ 
		F34 &  &  &  &  &  & 2 & 5 & 6 & 7 & 19 & \textbf{24} & 18 & 7 & 3 & 2 & 7 & 2 &   & \textit{102} \\ 
		F35 &  & 1 &  &  &  &  & 1 &  &  &  & 1 & 1 &  &  & 1 &  & 1 &   & \textit{6} \\ 
		F36 &  &  &  &  &  &  &  &  &  &  & 1 & 1 & 1 &  &  &  &  &   & \textit{3} \\ 
		F37 &  &  &  &  & 1 & 1 &  & 7 & 4 & 15 & 22 & 36 & \textbf{43} & 12 & 15 & 5 &  &   & \textit{161} \\ 
		F38 &  &  &  &  &  &  &  &  &  & 5 & 1 & 3 & 7 & 4 & 5 & 7 & 7 &   & \textit{39} \\ 
		F39 &  &  & 1 &  &  &  & 2 & 3 & 11 & 19 & 35 & 26 & \textbf{75} & 31 & 14 & 18 & 3 & 6  & \textit{244} \\ 
		F40 &  &  &  &  &  &  &  &  &  & 9 & 13 & 17 & 26 & \textbf{31} & 21 & 4 & 5 &   & \textit{126} \\ 
		F41 &  &  &  & 2 & 1 &  &  &  &  & 1 & 12 & 17 & 44 & \textbf{55} & 17 & 4 & 5 &   & \textit{158} \\ 
		F42 &  &  &  &  &  &  &  & 2 & 3 & 4 & 19 & 12 & 31 & 29 & \textbf{37} & 22 & 11 & 2  & \textit{172} \\ 
		F43 &  &  &  &  &  &  &  &  & 1 & 3 & 7 & 7 & 30 & 26 & 29 & \textbf{33} & 20 & 12  & \textit{168} \\ 
		F44 &  &  &  &  &  &  &  &  &  & 1 &  & 5 & 15 & 26 & \textbf{72} & 46 & 40 & 15  & \textit{220} \\ 
		F45 &  &  &  & 1 &  &  &  & 1 &  & 2 & 2 & 2 & 2 & 1 & 1 & \textbf{3} & 1 &   & \textit{16} \\ 
		F46 &  &  &  &  &  &  &  &  &  &  & 5 & 2 & 3 & 27 & 89 & 46 & \textbf{129} & 3  & \textit{304} \\ 
		F47 & 2 & 1 &  &  &  & 1 &  &  & 1 & 3 & 2 & 3 & 2 &  &  & 1 & \textbf{4} &   & \textit{20} \\ 
	\textit{Total} & \textit{18}  & \textit{227}  & \textit{44}  & \textit{189}  & \textit{196}  & \textit{152}  & \textit{196}  & \textit{257}  & \textit{219}  & \textit{517}  & \textit{472}  & \textit{303}  & \textit{448}  & \textit{285}  & \textit{355}  & \textit{263}  & \textit{248}  & \textit{69}   & \textit{4458}  \\ 
		\hline
	\end{tabular}
\end{table*}

\begin{figure}
	\includegraphics[width=\columnwidth]{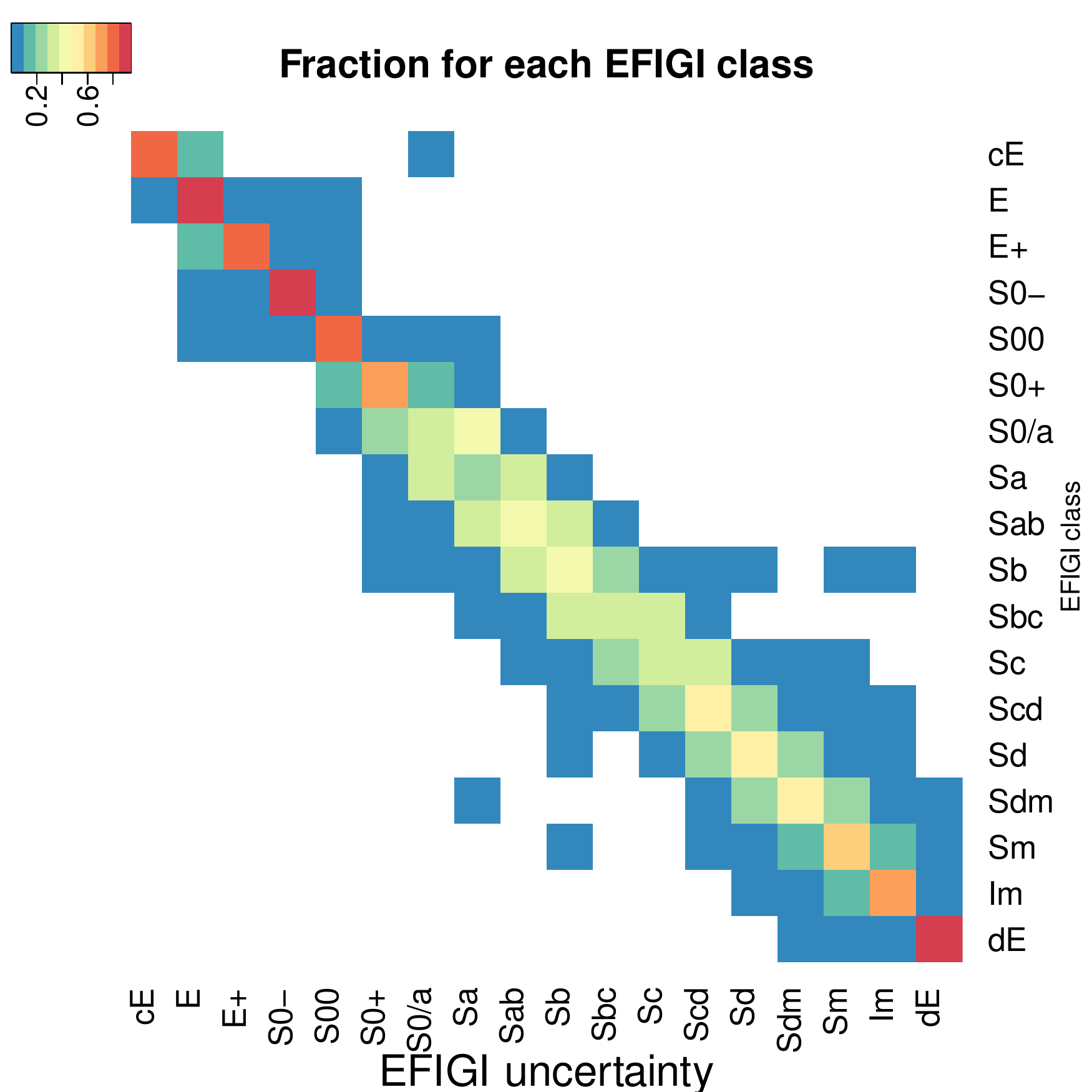}
		\caption{Uncertainty of the EFIGI classification. The vertical axis gives the morphology class given by the EFIGI catalogue, and the horizontal axis gives the morphology {types} attributed by all the astronomers. Hence, the sum of the fraction given by the colour code along the horizontal lines is 1.}
	\label{fig:heatmapTunc}
\end{figure}

\begin{figure*}
	\includegraphics[width=\columnwidth]{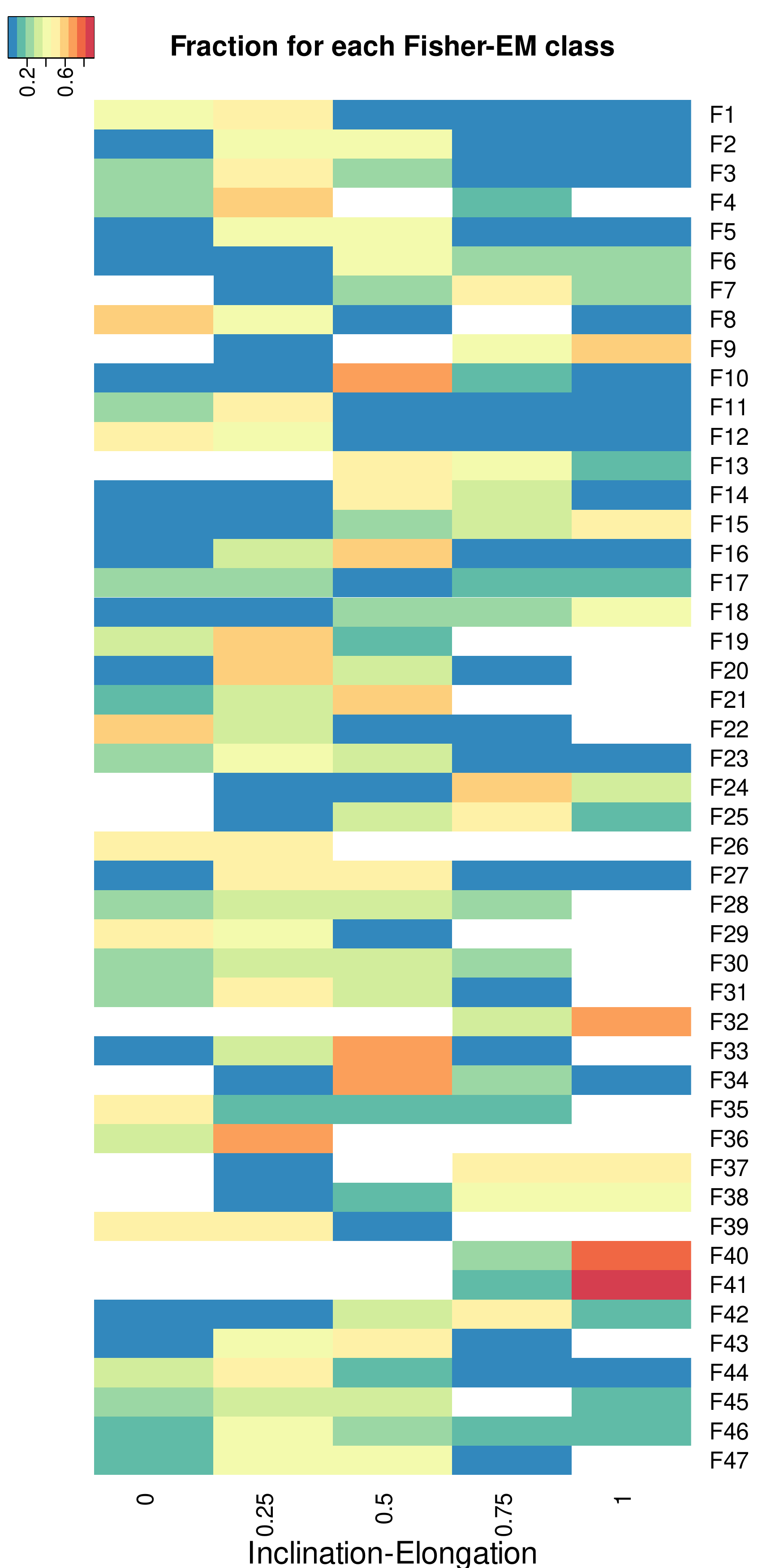}
    \includegraphics[width=\columnwidth]{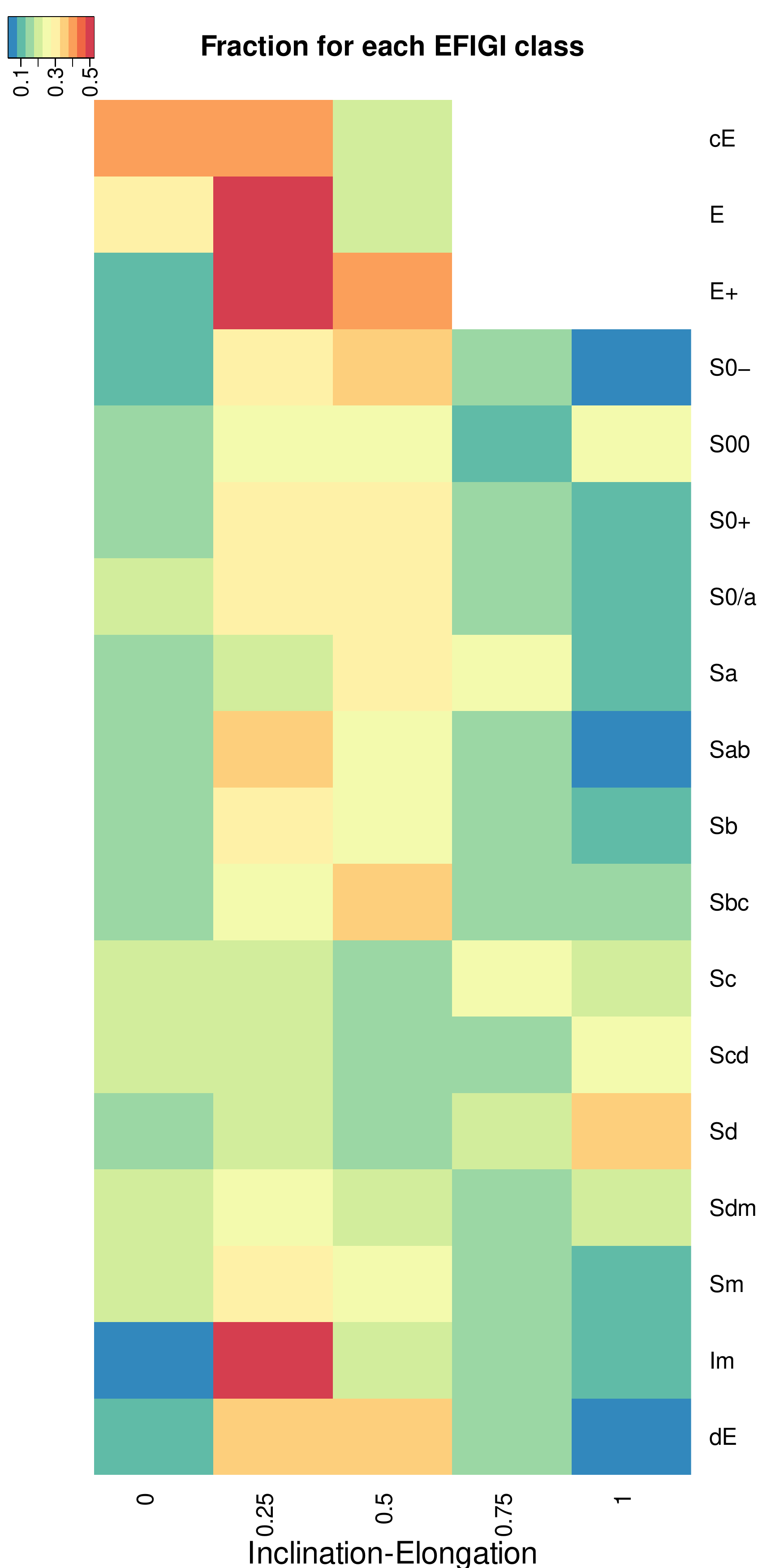}
	\caption{Heatmap of the Inclination-Elongation EFIGI parameter for the Fisher-EM classification (left) and the EFIGI classification (right). Note that the colours codes are different, going from 0.1 to 1 on the left diagram and from 0.05 to 0.5 on the right diagram.}
	\label{fig:heatmapInclination}
\end{figure*}

\begin{figure*}
	\includegraphics[width=\columnwidth]{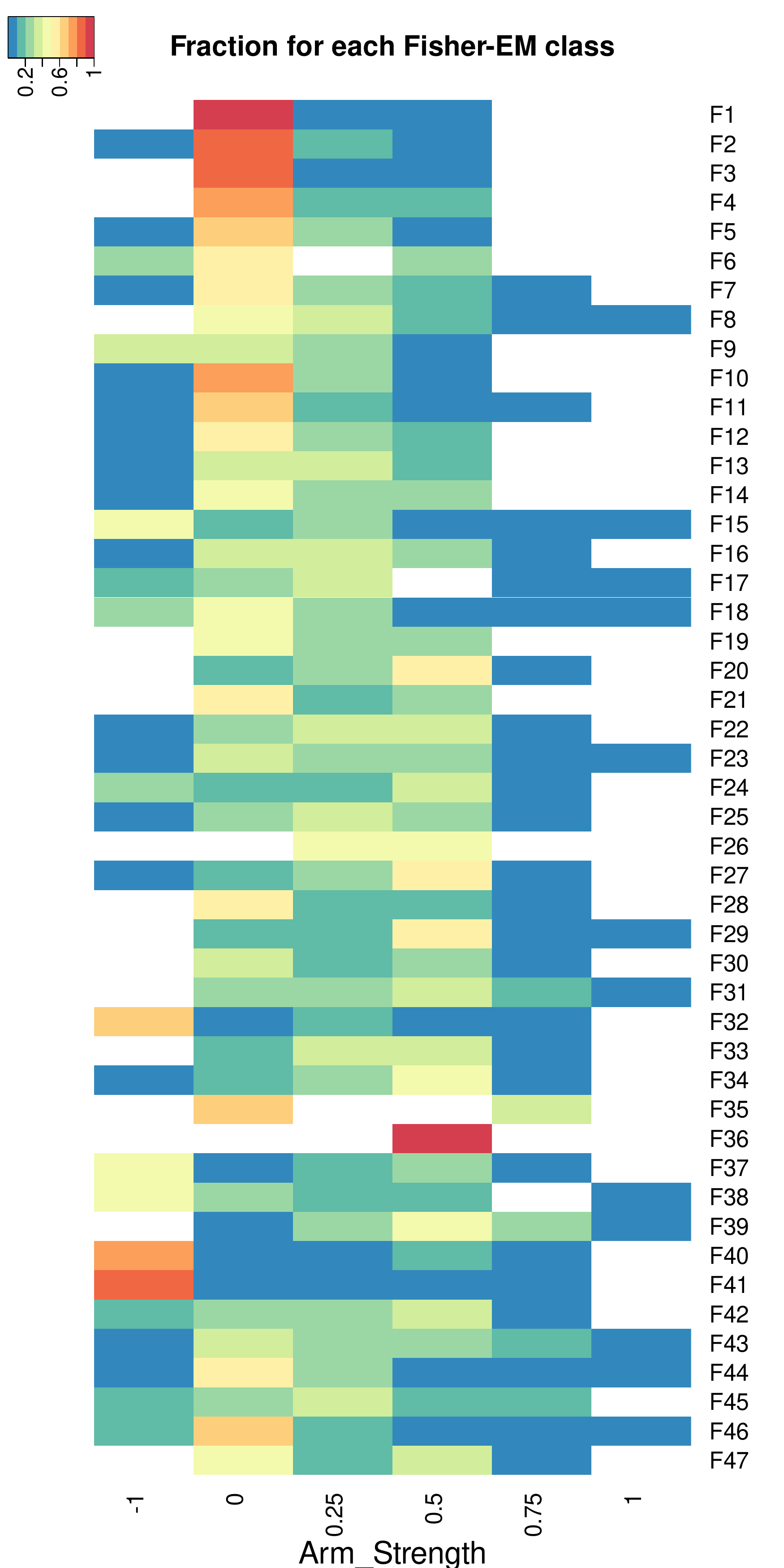}
    \includegraphics[width=\columnwidth]{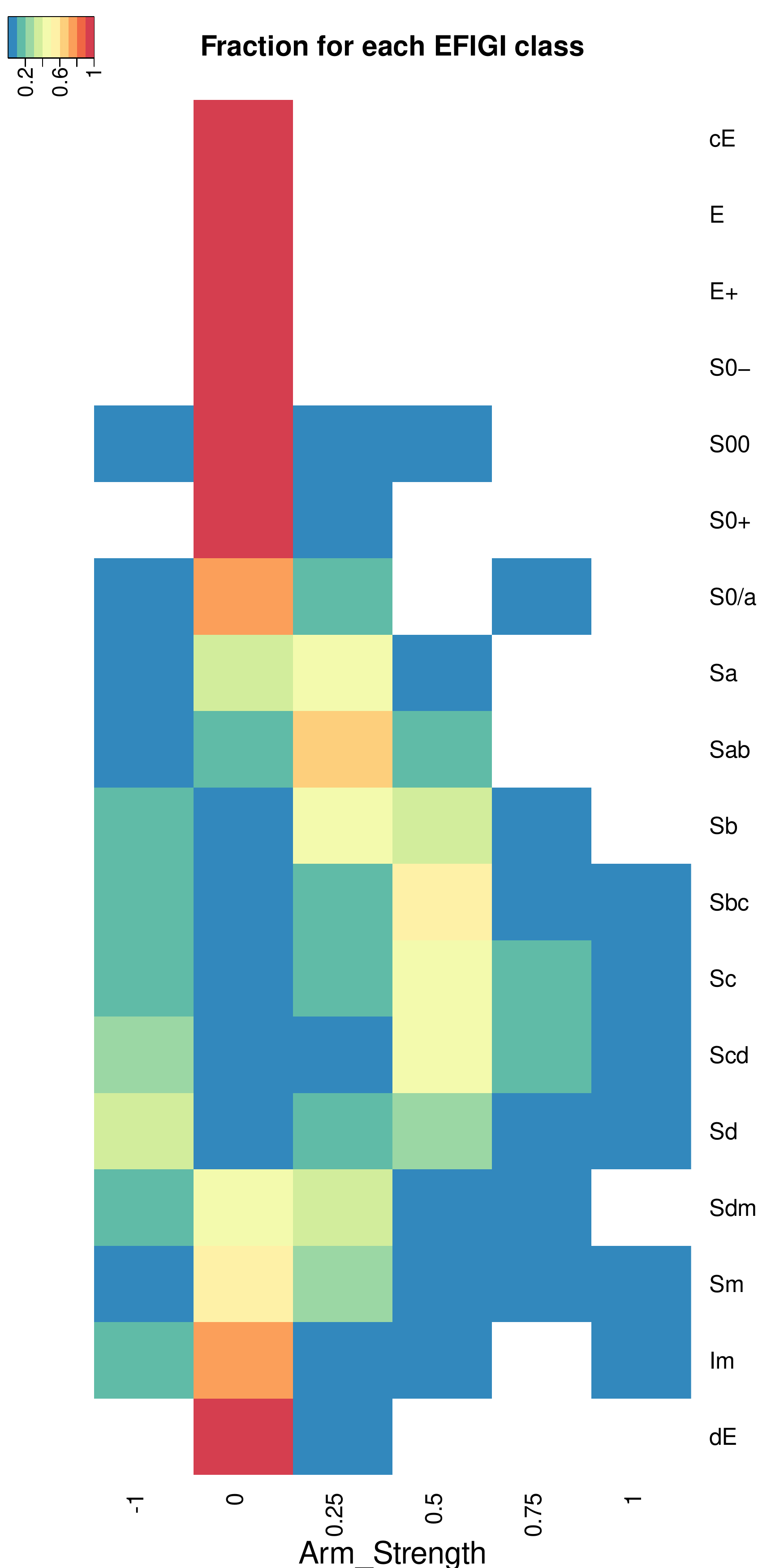}
	\caption{Heatmap of the Arm Strength EFIGI parameter for the Fisher-EM classification (left) and the EFIGI classification (right). }
	\label{fig:heatmapArmStrength}
\end{figure*}

\begin{figure*}
	\includegraphics[width=\textwidth]{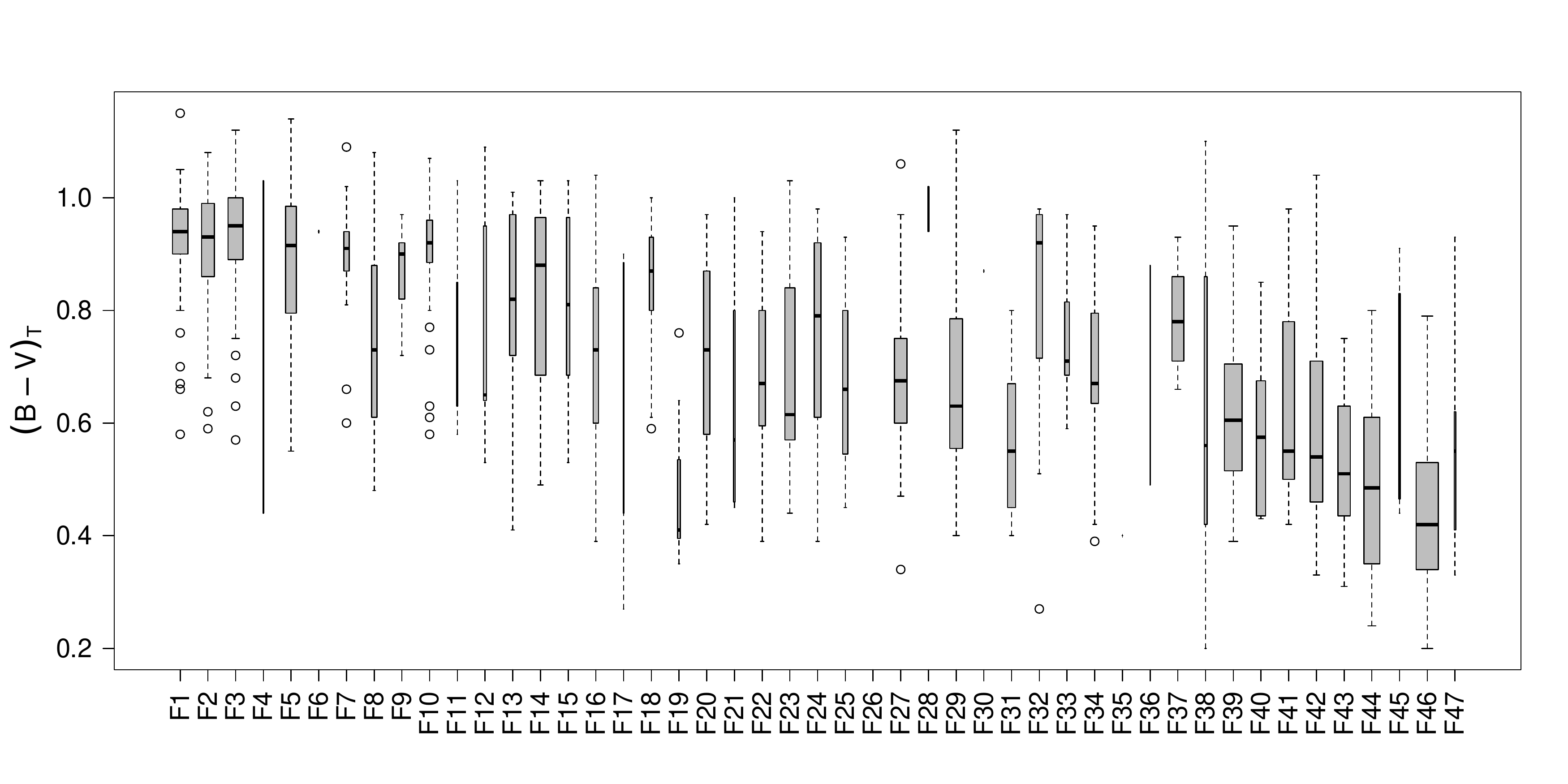} 
	\includegraphics[width=\textwidth]{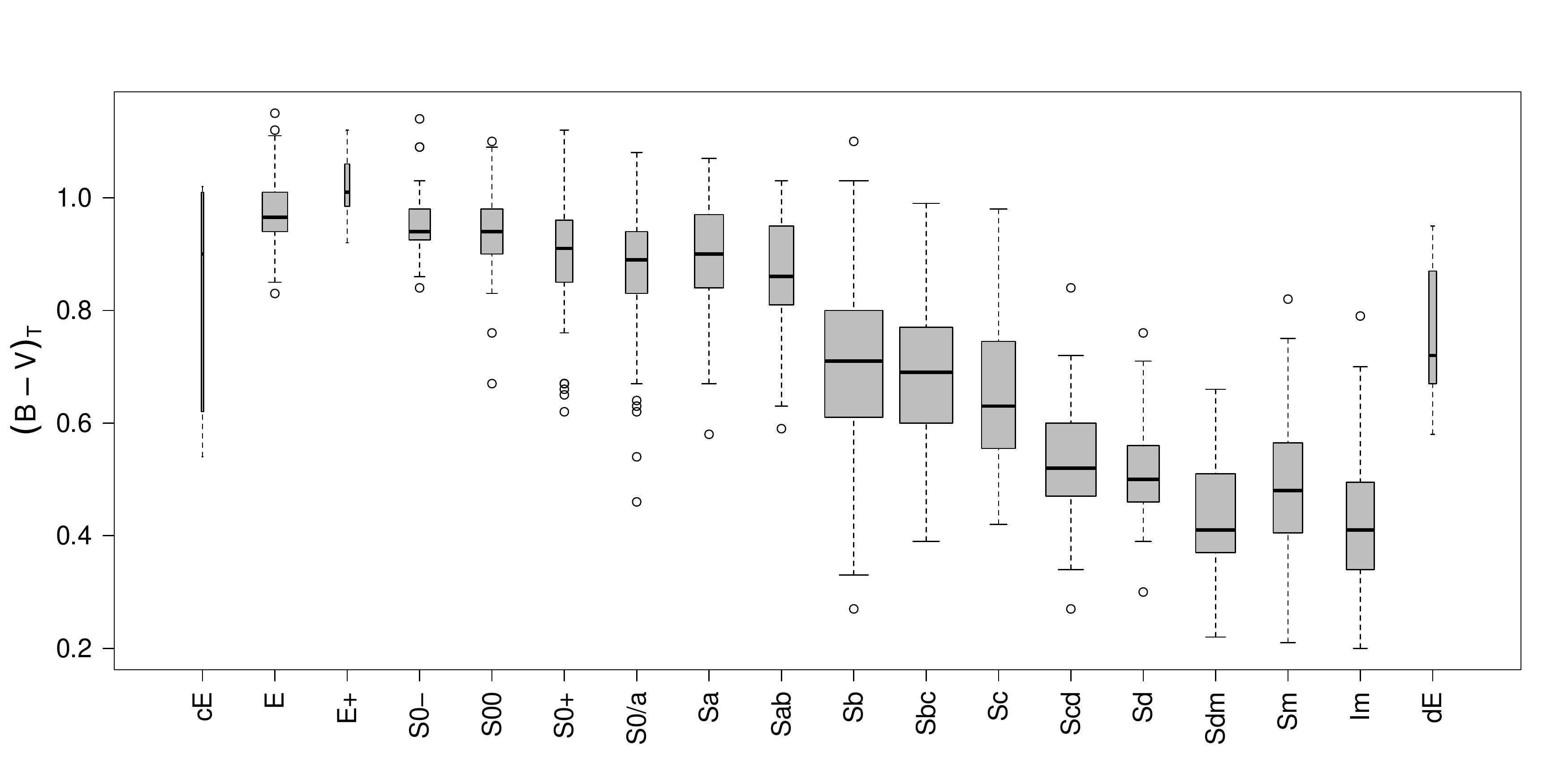}
	\caption{Boxplot of the color $(B-V)_T$ for the Fisher-EM classification (top) and the EFIGI classification (bottom). The height of the boxes show the quartiles, and their width is proportional to the size of the class.}
	\label{fig:heatmapBmV}
\end{figure*}

\begin{figure*}
	\includegraphics[width=\textwidth]{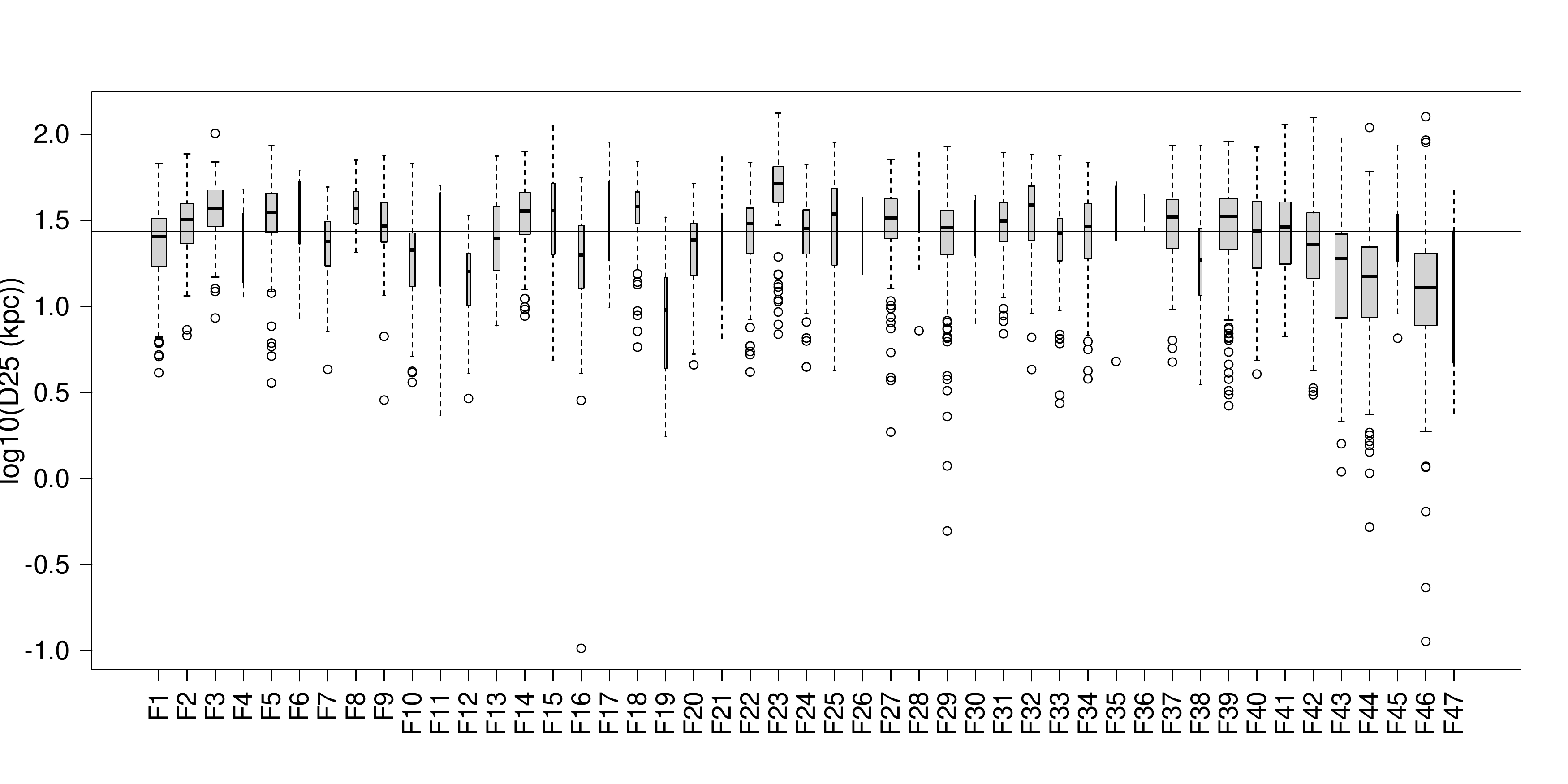} 
	\includegraphics[width=\textwidth]{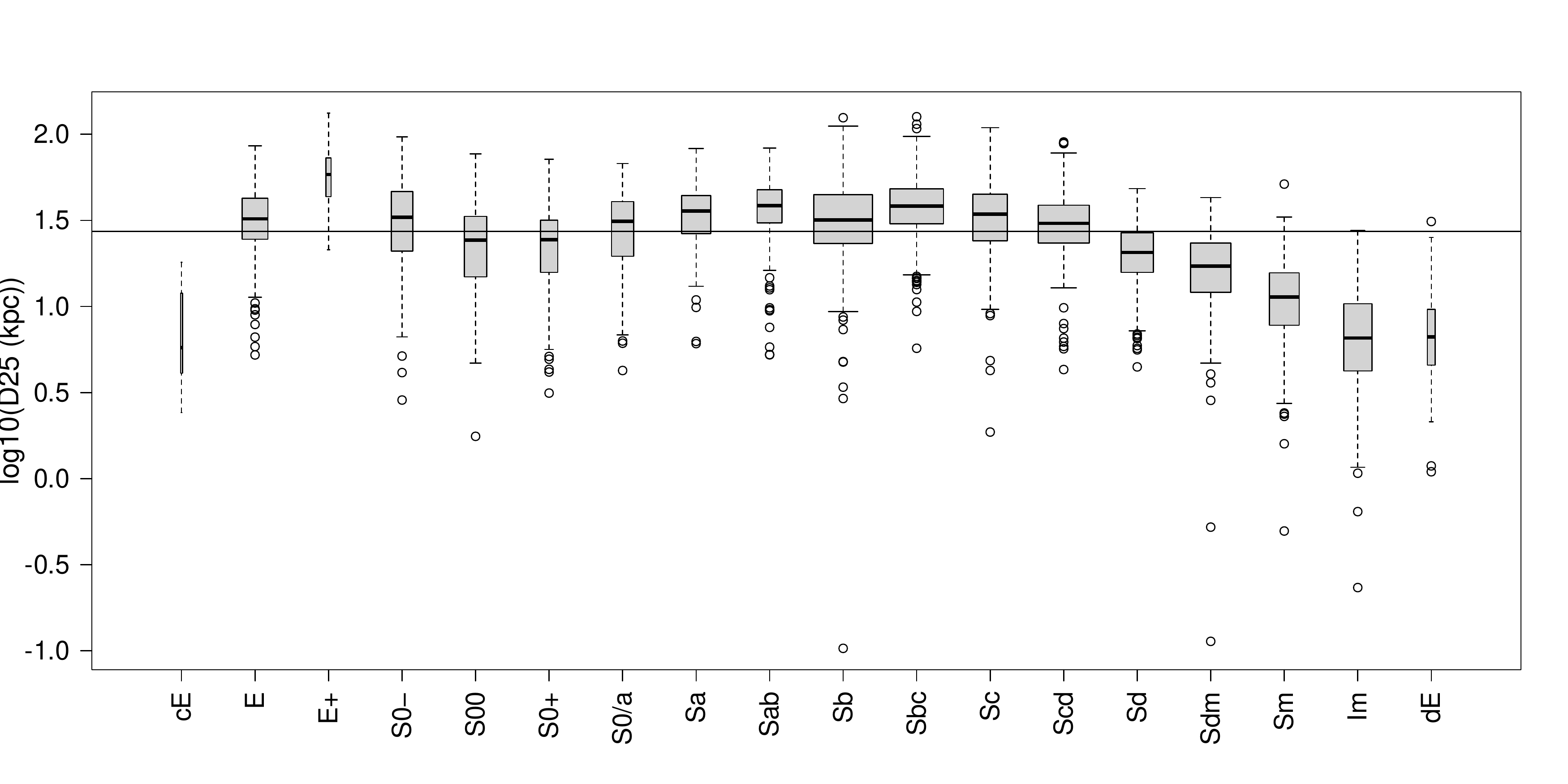}
	\caption{Boxplot of the diameter D25 for the Fisher-EM classification (top) and the EFIGI classification (bottom). The height of the boxes show the quartiles, and their width is proportional to the size of the class. The horizontal line is the median value for the whole sample.}
	\label{fig:heatmapD25kpc}
\end{figure*}

\begin{figure*}
	\includegraphics[width=\columnwidth]{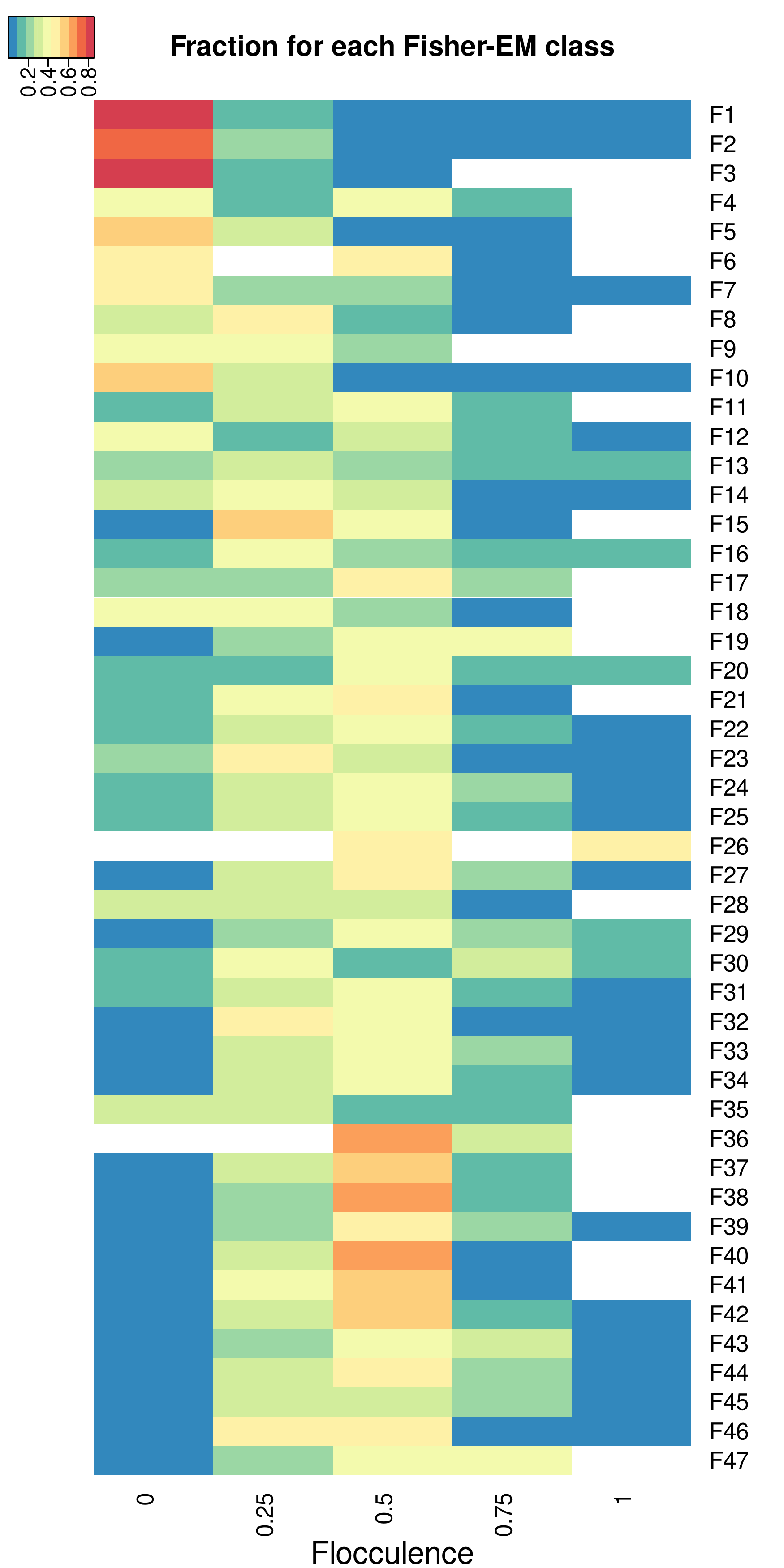}
	\includegraphics[width=\columnwidth]{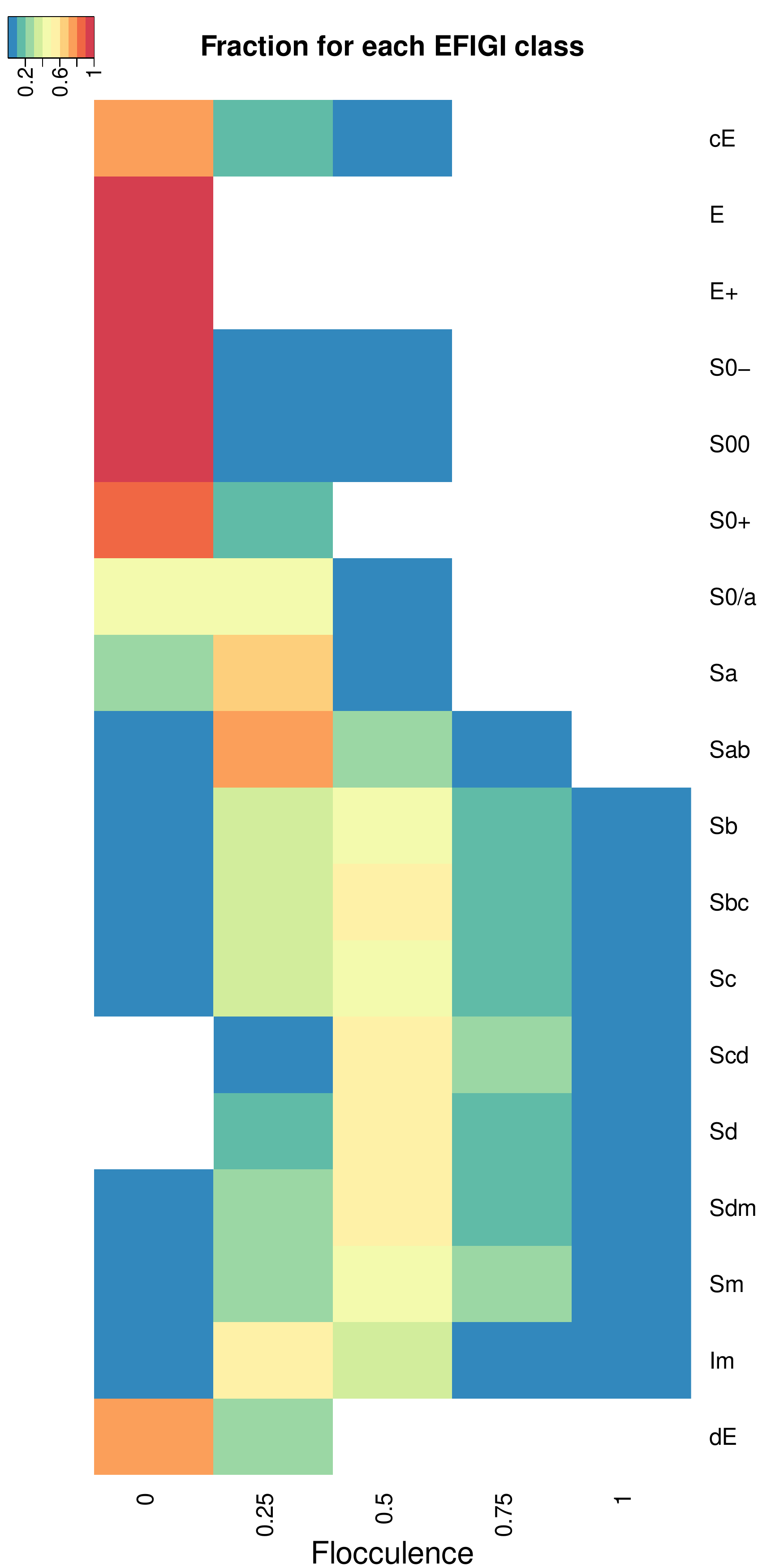}
	\caption{Heatmap of the Flocculence EFIGI parameter for the Fisher-EM classification (left) and the EFIGI classification (right). Note that the colour codes are not exactly the same (see inlets).}
	\label{fig:heatmapFlocculence}
\end{figure*}

\begin{figure*}
	\includegraphics[width=\columnwidth]{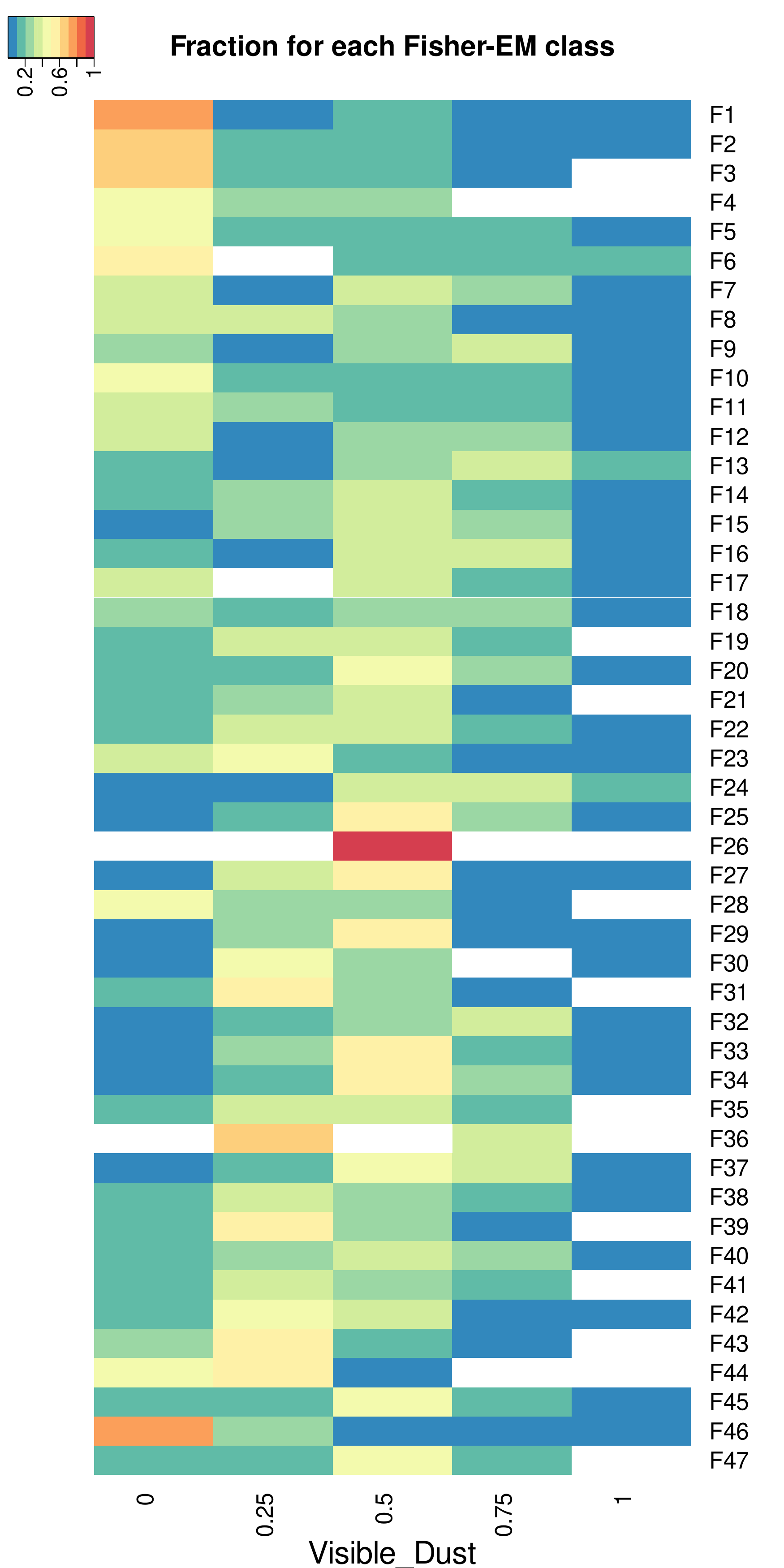}
	\includegraphics[width=\columnwidth]{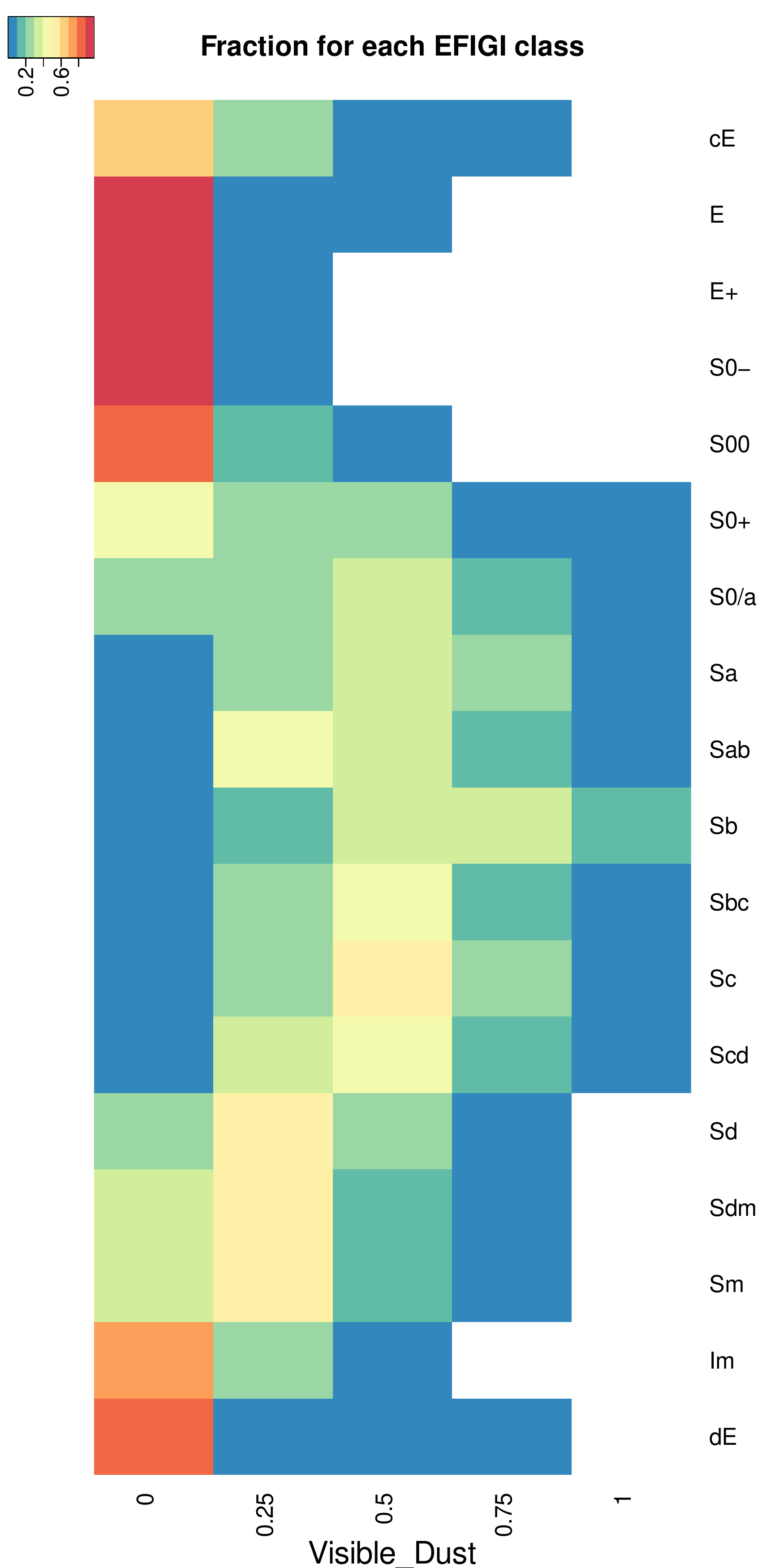}
	\caption{Heatmap of the Visible Dust EFIGI parameter for the Fisher-EM classification (left) and the EFIGI classification (right). Note that the colour codes are not exactly the same (see inlets).}
	\label{fig:heatmapVisibleDust}
\end{figure*}

\bsp	
\label{lastpage}
\end{document}